\documentclass[longauth]{aa}
\usepackage[utf8]{inputenc}
\usepackage[colorlinks,allcolors=blue]{hyperref}
\makeatletter
\renewcommand*\aa@pageof{, page \thepage{} of \pageref*{LastPage}}
\makeatother

\title{Spectral and orbital characterisation of the directly imaged giant planet HIP\,65426\,b\thanks{Based on observations collected at the European Organisation for Astronomical Research in the Southern Hemisphere under ESO programmes 199.C-0065 (PI: Launhardt), 198.C-0209 (PI: Beuzit) and 1100.C-0481 (PI: Beuzit)}}

\titlerunning{Spectral and orbital characterization of HIP\,65426\,b}
\author{
		A. C. Cheetham\inst{\ref{inst:Geneva}} \and 
		M. Samland\inst{\ref{inst:MPIA}} \and 
		S. S. Brems\inst{\ref{inst:LSW}} \and
		R. Launhardt\inst{\ref{inst:MPIA}} \and
		G. Chauvin\inst{\ref{inst:IPAG},\ref{inst:UMIFCA}} \and
		D. S{\'e}gransan\inst{\ref{inst:Geneva}} \and
		T. Henning\inst{\ref{inst:MPIA}} \and 
		A. Quirrenbach\inst{\ref{inst:LSW}} \and
		H. Avenhaus\inst{\ref{inst:MPIA}} \and 
		G. Cugno\inst{\ref{inst:ETHZ}} \and
		J. Girard\inst{\ref{inst:STScI}} \and
		N. Godoy\inst{\ref{inst:Valparaiso},\ref{inst:NPF}} \and
		G. M. Kennedy\inst{\ref{inst:Warwick}} \and
		A.-L. Maire\inst{\ref{inst:MPIA}} \and 
		S. Metchev\inst{\ref{inst:LSW},\ref{inst:UWO},\ref{inst:SBU}} \and
		A. Müller\inst{\ref{inst:MPIA}} \and 
		A. Musso Barcucci\inst{\ref{inst:MPIA}} \and 
		J. Olofsson\inst{\ref{inst:Valparaiso},\ref{inst:NPF}} \and
		F. Pepe\inst{\ref{inst:Geneva}} \and
		S. P. Quanz\inst{\ref{inst:ETHZ}} \and
		D. Queloz\inst{\ref{inst:Cambridge}} \and
		S. Reffert\inst{\ref{inst:LSW}} \and
		E. Rickman\inst{\ref{inst:Geneva}} \and
		R. van Boekel\inst{\ref{inst:MPIA}} \and 
		A. Boccaletti\inst{\ref{inst:LESIA}} \and
		M. Bonnefoy\inst{\ref{inst:IPAG}} \and
		F. Cantalloube\inst{\ref{inst:MPIA}} \and 
		B. Charnay\inst{\ref{inst:LESIA}} \and
		P. Delorme\inst{\ref{inst:IPAG}} \and
		M. Janson\inst{\ref{inst:MPIA},\ref{inst:Stockholm}} \and
		M. Keppler\inst{\ref{inst:MPIA}} \and
		A.-M. Lagrange\inst{\ref{inst:IPAG}} \and
		M. Langlois\inst{\ref{inst:LAM},\ref{inst:Lyon}} \and
		C. Lazzoni\inst{\ref{inst:INAF}} \and
		F. Menard\inst{\ref{inst:IPAG}} \and
		D. Mesa\inst{\ref{inst:INAF},\ref{inst:INCT}} \and
		M. Meyer\inst{\ref{inst:Michigan}} \and
		T. Schmidt\inst{\ref{inst:LESIA}} \and
		E. Sissa\inst{\ref{inst:INAF}} \and
		S. Udry\inst{\ref{inst:Geneva}} \and
		A. Zurlo\inst{\ref{inst:LAM},\ref{inst:UDP_chile},\ref{inst:UDP_chile2}}
		}
\institute{
		D{\'e}partement d'Astronomie, Universit{\'e} de Gen{\`e}ve, 51 chemin des Maillettes, 1290, Versoix, Switzerland \email{anthony.cheetham@unige.ch} \label{inst:Geneva} \and
		Max Planck Institute for Astronomy, K{\"o}nigstuhl 17, D-69117 Heidelberg, Germany\label{inst:MPIA} \and
		Landessternwarte, Zentrum f{\"u}r Astronomie der Universit{\"a}t Heidelberg, K{\"o}nigstuhl 12, D-69117 Heidelberg, Germany\label{inst:LSW} \and
		Universit{\'e} Grenoble Alpes, CNRS, IPAG, 38000 Grenoble, France\label{inst:IPAG} \and
		Unidad Mixta Internacional Franco-Chilena de Astronom{\'i}a,CNRS/INSU UMI 3386 and Departamento de Astronom{\'i}a, Universidad de Chile, Casilla 36-D, Santiago, Chile\label{inst:UMIFCA} \and
		Institute for Particle Physics and Astrophysics, ETH Zurich, Wolfgang-Pauli-Strasse 27, 8093 Zurich, Switzerland\label{inst:ETHZ} \and
		Space Telescope Science Institute, 3700 San Martin Dr. Baltimore, MD 21218, USA\label{inst:STScI} \and
		Instituto de F{\'i}sica y Astronom{\'i}a, Facultad de Ciencias, Universidad de Valpara{\'i}so, Av. Gran Bretaña 1111, Playa Ancha, Valpara\'{i}so, Chile\label{inst:Valparaiso} \and
		N{\'u}cleo Milenio Formaci{\'o}n Planetaria - NPF, Universidad de Valpara{\'i}so, Av. Gran Bretaña 1111, Valpara{\'i}so, Chile\label{inst:NPF} \and
		Department of Physics, University of Warwick, Gibbet Hill Road, Coventry, CV4 7AL, UK\label{inst:Warwick} \and
		Department of Physics and Astronomy, Centre for Planetary Science and Exploration, The University of Western Ontario, London, ON N6A 3K7, Canada\label{inst:UWO} \and
		Department of Physics and Astronomy, Stony Brook University, Stony Brook, NY 11794-3800, USA\label{inst:SBU} \and
		Astrophysics Group, Cavendish Laboratory, J.J. Thomson Avenue, Cambridge CB3 0HE, UK\label{inst:Cambridge} \and
	    LESIA, Observatoire de Paris, PSL Research University, CNRS, Sorbonne Universités, UPMC Univ. Paris 06, Univ. Paris Diderot, Sorbonne Paris Cité, 5 place Jules Janssen, 92195 Meudon, France\label{inst:LESIA} \and
	    Department of Astronomy, Stockholm University, AlbaNova University Center, SE-10691, Stockholm, Sweden\label{inst:Stockholm} \and
    	Aix Marseille Universit{\'e}, CNRS, LAM (Laboratoire d’Astrophysique de Marseille) UMR 7326, 13388 Marseille, France\label{inst:LAM} \and
    	CRAL, UMR 5574, CNRS, Universit{\'e} de Lyon, Ecole Normale Sup{\'e}rieure de Lyon, 46 Alle d’Italie, F-69364 Lyon Cedex 07, France\label{inst:Lyon} \and
    	INAF - Osservatorio Astronomico di Padova, Vicolo dell’ Osservatorio 5, 35122, Padova, Italy\label{inst:INAF} \and
    	INCT, Universidad De Atacama, calle Copayapu 485, Copiap\'{o}, Atacama, Chile\label{inst:INCT} \and
    	Department of Astronomy, University of Michigan, 1085 S. University Ave, Ann Arbor, MI 48109-1107, USA \label{inst:Michigan} \and
    	N\'ucleo de Astronom\'ia, Facultad de Ingenier\'ia y Ciencias, Universidad Diego Portales, Av. Ejercito 441, Santiago, Chile\label{inst:UDP_chile} \and
    	Escuela de Ingenier\'ia Industrial, Facultad de Ingenier\'ia y Ciencias, Universidad Diego Portales, Av. Ejercito 441, Santiago, Chile\label{inst:UDP_chile2}
		}
   
\abstract{HIP\,65426\,b is a recently discovered exoplanet imaged during the course of the SPHERE-SHINE survey. Here we present new $L'$ and $M'$ observations of the planet from the NACO instrument at the VLT from the NACO-ISPY survey, as well as a new $Y-H$ spectrum and $K$-band photometry from SPHERE-SHINE. Using these data, we confirm the nature of the companion as a warm, dusty planet with a mid-L spectral type. From comparison of its SED with the BT-Settl atmospheric models, we derive a best-fit effective temperature of $T_{\text{eff}}=1618\pm7$\,K, surface gravity $\log g=3.78^{+0.04}_{-0.03}$ and radius $R=1.17\pm0.04$\,$R_{\text{J}}$ (statistical uncertainties only). Using the DUSTY and COND isochrones we estimate a mass of $8\pm1$\,$M_{\text{J}}$.
Combining the astrometric measurements from our new datasets and from the literature, we show the first indications of orbital motion of the companion (2.6$\sigma$ significance) and derive preliminary orbital constraints. We find a highly inclined orbit ($i=107^{+13}_{-10}$\,deg) with an orbital period of $800^{+1200}_{-400}$\,yr.
We also report SPHERE sparse aperture masking observations that investigate the possibility that HIP\,65426\,b was scattered onto its current orbit by an additional companion at a smaller orbital separation. From this data we rule out the presence of brown dwarf companions with masses greater than 16\,$M_{\text{J}}$ at separations larger than 3\,AU, significantly narrowing the parameter space for such a companion.}

\begin{document}

\maketitle

\section{Introduction}

The number of exoplanets that can be studied through direct imaging is steadily growing \citep[e.g.][]{2015Sci...350...64M,2017A&A...605L...9C,2018arXiv180611568K}, and each object can provide a wealth of information about their formation, evolution and atmospheres. 
For studies of giant planet atmospheres, the mid-infrared wavelength range covers a critical regime that is particularly sensitive to cloud properties, and contains strong molecular absorption bands from CH$_4$ and CO \citep{2007ApJS..168..140S}. Since young giant planets still retain the heat of their formation, this wavelength range also offers favourable flux ratios between planet and host star. This has led to the majority of the directly imaged planets being discovered or studied in this wavelength range, particularly in the $L$-band \citep[e.g.][]{2004A&A...425L..29C,2008Sci...322.1348M,2010Sci...329...57L,2010Natur.468.1080M,2013ApJ...772L..15R,2014ApJ...792...17S,2015Sci...350...64M,2018arXiv180611567M}. However, the increasing thermal background at longer wavelengths makes detections beyond the $L$-band much more difficult and only a few planets have published $M$-band photometry, including HR 8799 b,c,d \citep{2011ApJ...739L..41G}, $\beta$ Pic b \citep{2013A&A...555A.107B} and 51 Eri b \citep{2017AJ....154...10R}. JWST will be a critical tool for studying such objects at mid-infrared wavelengths, with the NIRCAM and MIRI instruments having high-contrast coronagraphic imaging capabilities across 1.8\,$\mu$m to 25\,$\mu$m \citep{2007SPIE.6693E..0HK,2015PASP..127..633B}.

One planet that has yet to be studied in the mid-infrared is HIP\,65426\,b, recently discovered by the SHINE (SpHere INfrared survey for Exoplanets) survey utilising the SPHERE instrument at the VLT \citep{2017sf2a.conf..331C}. Analysis of its 1-2.3$\mu$m spectrum shows a likely L5-L7 spectral type, effective temperature of $T_\text{eff}=1300-1600$\,K, low surface gravity and a luminosity consistent with a 6-12\,$M_{\text{J}}$ planet. Given its placement in the mid-L spectral sequence, it provides an important opportunity to study the atmospheric physics of young giant planets in a regime where complex cloud/dust physics and non-equilibrium chemistry play key roles \citep[e.g.][]{2011ApJ...739L..41G,2014ApJ...792...17S,2014ApJ...795..133C}. This object is also a primary target of a JWST Early Release Science program \citep{2017jwst.prop.1386H}, which will provide additional photometry in the near- and mid-infrared.

In this paper, we present the results of new mid-infrared imaging taken during the ISPY survey (Imaging Survey for Planets around Young stars; Launhardt et al. 2018, in prep) with the NACO instrument at the VLT, as well as new near-infrared coronagraphic and Sparse Aperture Masking (SAM) observations obtained with SPHERE. We adopt the stellar parameters of HIP\,65426 from \cite{2017A&A...605L...9C}. In particular we assume a solar metallicity, an effective temperature of $8840\pm200$\,K estimated from HARPS spectroscopy, as well as a primary mass of $1.96\pm0.04$\,$M_\odot$ and age of $14\pm4$\,Myr from comparison of isochrones with the photometry of HIP\,65426 and that of neighbouring stars. However, we update its parallax and distance using the results of the Gaia DR2 \citep{2016A&A...595A...1G,2018arXiv180409365G}. This yields a distance of $109.2\pm0.7$\,pc \citep[c.f. $111.4\pm3.8$ pc from][]{2017A&A...605L...9C}, and an updated projected separation for the companion of $90$\,AU.

\section{Observations and Data Analysis}
\subsection{NACO imaging}
HIP\,65426 was observed as part of the NACO-ISPY survey (Launhardt et al. 2018, in prep) on 2017-05-18 and 2017-05-19 using the $L'$ ($3.8\mu$m) and $M'$ ($4.8\mu$m) filters respectively of the NACO instrument \citep{1998SPIE.3354..606L,2003SPIE.4839..140R}. The majority of our observing time consisted of short exposure images with no coronagraph, dithering the star between 3 quadrants of the NACO detector\footnote{We avoided the detector quadrant with persistent striping present since the recommissioning of NACO in 2014-2015. This quadrant is in the lower-left, containing pixel (0,0).}. Despite the short integration time, the stellar PSF was partially saturated on the detector in $L'$. More details of the observing sequence can be found in Table \ref{tab:observing_log}. We also acquired a separate set of shorter exposure images to measure the stellar flux at the beginning and end of the sequence.

The data were reduced independently using the GRAPHIC \citep{2016MNRAS.455.2178H}, PynPoint \citep{2012MNRAS.427..948A}, and IPAG-ADI \citep{2012A&A...542A..41C} packages. These pipelines follow similar approaches for data cleaning. Similar results were obtained, and so we report the results of the GRAPHIC pipeline. Briefly, first the frames were sky subtracted using Principal Component Analysis (PCA) following the approach of \cite{2017arXiv170610069H}. The position of the star was then measured in each frame by fitting a Gaussian profile. We then compared the fitted parameters (amplitude, position and width) across the observing sequence and removed outliers more than 5 median absolute deviations from the median of each parameter. Frames were shifted to be centred on the star using Fourier transforms. The cleaned datacubes were then PSF (point spread function) subtracted using PCA \citep{2012ApJ...755L..28S,2012MNRAS.427..948A}. The cleaning procedure was also performed on the unsaturated, shorter exposure images to measure the stellar flux and as a reference for the unsaturated PSF shape.

\begin{figure}
\includegraphics[width=0.5\textwidth]{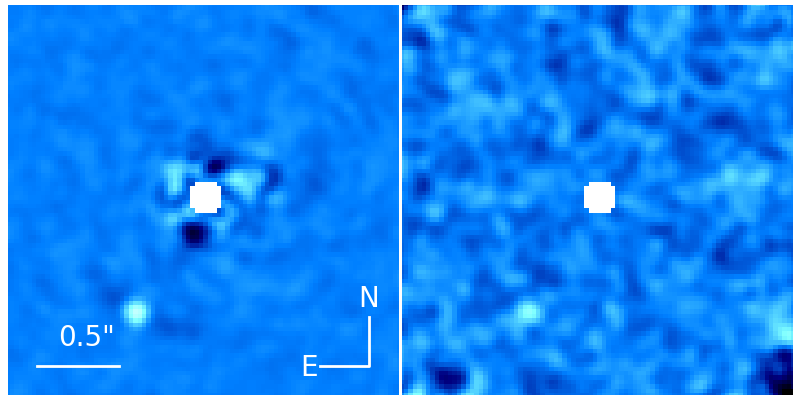}
\caption{Final PSF-subtracted images from the two NACO datasets in $L'$ (left) and $M'$ (right). The companion is detected in both datasets. While we find a lower contrast ratio in $M'$, the high thermal background leads to a lower overall SNR. }\label{fig:naco_images}
\end{figure}

For the GRAPHIC reduction, the data were first binned in groups of 126 frames for the $L'$ data and groups of 200 frames for the $M'$ data. PCA was applied to annular sections of the image, each with a width of 2 FWHM (full width at half maximum, measured from the stellar PSF). For each frame, a reference library was constructed using only those frames where the field rotation was large enough for a companion to have moved by 0.75 FWHM. 30\% of the available PCA modes were then subtracted from the data. The PSF-subtracted frames were then derotated using Fourier transforms \citep{larkin1997fast} and median-combined to produce the final images in Figure \ref{fig:naco_images}.

To extract the astrometry and photometry of the companion, we used the negative PSF injection technique \citep{2010Sci...329...57L}. For each set of parameters, we inject a negative copy of the unsaturated PSF into the raw frames with the appropriate position and flux. We then calculate the likelihood from the residuals in a small (10 pixel) box around the companion position. We assume Gaussian distributed residuals with a standard deviation measured from the values in an annulus at the same separation taken from the initial reduction, after masking out the companion. The same values were used to verify the Gaussianity with a one-sided Kolmogorov-Smirnov test (D=0.022, p=0.45 and D=0.019, p=0.56 for $L'$ and $M'$ respectively). The likelihood was measured on a grid of parameters around the best-fit position, and we extracted the uncertainties from the marginal likelihood distribution of each parameter.

In addition to the uncertainties resulting from the fit, we explored the effect of different PCA parameters, varying the protection angle between 0.5-1.5 FWHM and the number of PCA modes from 10\% to 40\% of the number of binned frames. The scatter of the best-fit values was added in quadrature to our astrometric and photometric uncertainties. We also added an additional term to the photometric uncertainty (13\% in $L'$ and 3\% in $M'$) due to the change in flux of the primary star measured during the short-exposure sequence at the start and end of the observing sequence, caused by variations in the observing conditions and adaptive optics correction. The larger variation seen in $L'$ was caused by the presence of thin clouds during the sequence.

To verify the robustness of the uncertainties calculated from this approach, we injected synthetic companions using the PSF of another star observed during the same night. First, we used the best-fit parameters of HIP\,65426\,b to remove the companion from the data and generate a clean datacube. We then injected companions with randomly drawn parameters, using fluxes and separations centred on that of HIP\,65426\,b, and with random position angles. For each iteration we performed the same procedure to measure the parameters of the injected companion. This was repeated 50 times for each filter. We then calculated the difference between the injected and measured parameters. The root-mean-square of these residuals were within 10\% of our uncertainties for HIP\,65426\,b, and so we conclude that they are representative of our measurement accuracy.

\begin{table*} 
\caption{Observing Log}  
\label{tab:observing_log}  
\center
\begin{tabular}{cccccccc}
\hline
UT Date & Instrument & Filter & NDIT$^{a}\times$DIT$^{b}$ & NDIT$^{a}\times$DIT$^{b}$ (Flux$^{c}$) & $\Delta \pi ^{d}$ & True North & Pixel Scale \\
& & & [s] & [s] & [$^{\circ}$] & [$^{\circ}$] & [mas/pixel] \\
\hline
2017-05-18 & NACO  & $L'$ & $30114\times0.2$ & $2400\times0.1$ & 113 & $-0.44\pm0.10$ & $27.20\pm0.05$ \\
2017-05-19 & NACO  & $M'$ & $71200\times0.05$ & $2400\times0.04$ & 96 & $-0.44\pm0.10$ & $27.20\pm0.05$ \\
2017-05-03 & SPHERE-IFS  & Y-H + SAM & $384\times4$ &  & 49.5 & $1.78\pm0.05$ & $7.46\pm0.02$ \\
2017-05-03 & SPHERE-IRDIS & K1 + SAM & $720\times2$ &  & 49.5 & $1.78\pm0.05$ & $12.267\pm0.01$ \\
2017-05-03 & SPHERE-IRDIS & K2 + SAM & $720\times2$ &  & 49.5 & $1.78\pm0.05$ & $12.263\pm0.01$ \\
2018-05-12 & SPHERE-IFS & Y-H & $40\times96$ & $47\times2$ & 31.7 & $1.76\pm0.06$ & $7.46\pm0.02$ \\
2018-05-12 & SPHERE-IRDIS & K1 & $40\times96$ & $101\times0.84$ & 31.7 & $1.76\pm0.06$ & $12.267\pm0.01$ \\
2018-05-12 & SPHERE-IRDIS & K2 & $40\times96$ & $101\times0.84$ & 31.7 & $1.76\pm0.06$ & $12.263\pm0.01$ \\
\hline                  
\end{tabular}
\tablefoot{$^a$NDIT refers to the total number of integrations. $^b$DIT refers to the integration time of each image. $^{c}$Flux refers to the sequence of images taken to measure the number of detector counts and PSF shape of the primary star. $^d \Delta \pi$ is the maximum change in parallactic angle during the sequence.}
\end{table*}

To convert the contrast measurements of HIP\,65426\,b to physical fluxes we produced a synthetic spectrum by scaling a BT-NextGen spectrum \citep{2012RSPTA.370.2765A} with an effective temperature T$_\text{eff} = 8800$\,K, surface gravity $\log g =4.5$ and solar metallicity to the photometry of HIP\,65426 compiled from 2MASS, Tycho-2, HIPPARCOS and WISE \citep{skrutskie2006two,2000AA...355L..27H,2007AA...474..653V,2010AJ....140.1868W}. The contrast ratio measurements were then multiplied by the predicted flux of the primary star in each filter. We then used the Gaia DR2 distance of $109.2\pm0.7$\,pc \citep{2016A&A...595A...1G,2018arXiv180409365G} to convert to absolute flux, with the results given in Table \ref{tab:photometry}.

\subsection{SPHERE coronagraphic imaging}
As part of the SHINE exoplanet survey \citep{2017sf2a.conf..331C}, HIP\,65426 was re-observed on 2018-05-12 with deep coronagraphic imaging to improve the Y-H spectrum and K-band photometry of the companion, and to monitor its relative astrometry. We used the infrared dual-band imager and spectrograph \citep[IRDIS;][]{2008SPIE.7014E..3LD} and integral field spectrograph \citep[IFS;][]{2008SPIE.7014E..3EC} modules of SPHERE \citep{2008SPIE.7014E..18B}. Data were obtained using the IRDIFS\_EXT mode. In this configuration, IRDIS and IFS operate simultaneously, with IRDIS in dual-band imaging mode \citep{2010MNRAS.407...71V} using the K1 and K2 filters ($2.100\mu$m and $2.251\mu$m respectively) and IFS covering the Y-H bands ($0.96-1.64\mu$m).

The data processing follows the same procedure as in \cite{2017A&A...605L...9C}. The data were first cleaned using the SPHERE Data Reduction and Handling (DRH) pipeline \citep{2008SPIE.7019E..39P}, consisting of background subtraction, flat fielding and extraction of the spectral data cube. Additional cleaning steps for IFS were performed using the routines described in \citet{2015A&A...576A.121M}, consisting of bad pixel correction, spectral cross-talk correction and an improved wavelength calibration procedure. 

The SpeCal pipeline \citep{2018A&A...615A..92G} was used to perform the starlight subtraction and companion analysis. The TLOCI \citep{2014IAUS..299...48M} algorithm was used to perform the starlight subtraction, and we used the PSF model approach described in \citet{2018A&A...615A..92G} to measure the companion position and flux in each filter.

\subsection{SPHERE Aperture Masking}
During the SHINE survey, HIP\,65426 was also observed with the sparse aperture mask (SAM) mode of SPHERE \citep{2016SPIE.9907E..2TC} on 2017-05-03, using a 7-hole mask. This mode turns the telescope pupil into an interferometric array, allowing the use of observables such as the closure phase \citep{1958MNRAS.118..276J} that are robust to optical aberrations. This results in SAM being a useful technique for studying high-contrast structure or companions at small angular separations that would otherwise be inaccessible. While it was expected that HIP\,65426\,b would not be detectable with SAM due to the large separation and high contrast, the data were obtained to search for additional companions that would be undetectable with other techniques. The observations were taken with the IRDIFS\_EXT mode, the same filter combination as the coronagraphic imaging data. Further details of the observing sequence can be found in Table \ref{tab:observing_log}.

To clean the data, a different approach was taken for the IFS and IRDIS modules. The IFS data were cleaned using the same approach as the SPHERE coronagraphic data, while the IRDIS data were cleaned using a custom set of {\it IDL} routines consisting of background subtraction, flat fielding and bad pixel correction.

The cleaned datacubes were processed with a similar procedure to that outlined in \cite{tuthill2000keck} and \cite{kraus08}, with modifications made to deal with multi-wavelength data (39 channels for IFS and 2 for IRDIS). Briefly, images were centred and windowed with a super-Gaussian function to limit sensitivity to read-noise. Closure phases were then measured from the Fourier transforms of the images, and calibrated using a weighted sum of those measured on the reference star HD 116664, which was observed alternately with HIP\,65426 during the same sequence. Correlations between wavelengths (for both IFS and IRDIS) were not considered in this analysis, and so the detection limits are expected to be slightly optimistic.

To determine the limits for companion detection, a Monte-Carlo approach was taken. We generated 10,000 simulated closure phase datasets using the measured uncertainties. For each combination of separation, contrast and position angle on a grid of values, we counted the number of non-detections, defined as those simulations where a point source model fit the data better than the corresponding binary model. After marginalising over position angle, we calculated 3.3$\sigma$ detection limits for each separation as the contrast at which 99.9\% of noise simulations result in a non-detection. These were then converted to $5\sigma$ limits for consistency with the NACO and SPHERE imaging data.

\section{Results}

\begin{table*} 
\caption{Measured photometry of HIP\,65426\,b and derived masses}
\label{tab:photometry}
\center
\begin{tabular}{cccccc}
\hline
Instrument & Filter & $\Delta$mag & App. Mag & Abs. Mag & App. Flux [$Wm^{-2}\mu m^{-1}$]\\
\hline
SPHERE & $H_2$ & $11.14\pm0.05$ & $17.94\pm0.05$ & $12.75\pm0.05$ & $\left(8.6\pm0.4\right) \times 10^{-17} $ \\
SPHERE & $H_3$ & $10.78\pm0.06$ & $17.58\pm0.06$ & $12.39\pm0.06$ & $\left(10.1\pm0.6\right) \times 10^{-17} $ \\
SPHERE & $K_1$ & $10.19\pm0.10$ & $17.01\pm0.09$ & $11.82\pm0.09$ & $\left(7.5\pm0.6\right) \times 10^{-17} $ \\
SPHERE & $K_2$ & $9.82\pm0.10$ & $16.79\pm0.09$ & $11.60\pm0.09$ & $\left(7.1\pm0.6\right) \times 10^{-17} $ \\

NACO  & $L'$ & $8.47\pm0.14$ & $15.26\pm0.15$ & $10.07\pm0.15$ & $\left(4.1\pm0.5\right) \times 10^{-17} $ \\
NACO  & $M'$ & $8.2\pm0.4$ & $15.1\pm0.5$ & $9.9\pm0.5$ & $\left(2.1\pm0.8\right) \times 10^{-17} $ \\ 
\hline
\end{tabular}
\tablefoot{$^{a}$Best-fitting mass is at the edge of the model grid, and so is expressed as an upper limit.}
\end{table*}


\begin{table*} 
\caption{Measured astrometry of HIP\,65426\,b}
\label{tab:astrometry}
\center
\begin{tabular}{ccccccc}
\hline
UT Date & Instrument & Filter & $\rho$ [mas] & $\rho$ [AU] & $\theta$ [deg] & Reference\\
\hline
2016-05-30 & SPHERE & $H_2$ & $830.4\pm4.9$ & $90.7\pm0.8$ & $150.28\pm0.22$ & {\cite{2017A&A...605L...9C}} \\
2016-06-26 & SPHERE & $H_2$ & $830.1\pm3.2$ & $90.7\pm0.7$ & $150.14\pm0.17$ & {\cite{2017A&A...605L...9C}} \\
2017-02-07 & SPHERE & $H_2$ & $827.6\pm1.5$ & $90.4\pm0.6$ & $150.11\pm0.15$ & {\cite{2017A&A...605L...9C}} \\
2017-02-09 & SPHERE & $K_1$ & $828.8\pm1.5$ & $90.5\pm0.6$ & $150.05\pm0.16$ & {\cite{2017A&A...605L...9C}} \\
2017-05-18 & NACO  & $L'$ & $832\pm3$ & $90.8\pm0.7$ & $149.52\pm0.19$ & This work\\
2017-05-19 & NACO  & $M'$ & $850\pm20$ & $93\pm2$ & $148.5\pm1.6$ & This work \\
2018-05-12 & SPHERE & $K_1$ & $822.9\pm2.0$ & $89.9\pm0.6$ & $149.85\pm0.15$ & This work\\
2018-05-12 & SPHERE & $K_2$ & $826.4\pm2.4$ & $90.2\pm0.6$ & $149.89\pm0.16$ & This work\\

\hline                  
\end{tabular}
\end{table*}

\subsection{Astrometry}

To calibrate the NACO data, we compared $L'$ observations of an astrometric reference field in 47 Tuc taken on 2017-05-19 to the catalog of \citet{2006ApJS..166..249M}, resulting in the pixel scale and True North offset reported in Table \ref{tab:observing_log}. The final measured astrometry of HIP\,65426\,b is given in Table \ref{tab:astrometry}. The measured NACO $L'$ separation of $832\pm3$\,mas corresponds to a projected physical separation of $90.9\pm0.7$\,AU using the updated distance to HIP\,65426. For the SPHERE coronagraphic imaging data, we used the astrometric calibration scheme outlined in \cite{2016SPIE.9908E..34M}, with the pixel scale and True North offset also reported in Table \ref{tab:observing_log}.

Between the SPHERE epochs, we see the first indications of orbital motion, with the observed position changing consistently in the same direction and a maximum of $2.6\sigma$ deviation between the datapoints. To investigate the orbital parameters consistent with the observed motion, we used the Orbits for the Impatient (OFTI) method of \citet{2017AJ....153..229B}, based on Bayesian rejection sampling. We applied uniform priors in the epoch of periastron passage $T_0$ and argument of periastron $\omega$, a $\sin i$ prior in inclination $i$, and a linearly decreasing prior in the eccentricity $e$ based on the results of \citet{2008ApJ...674..466N}. To choose the semi-major axis $a$ and position angle of nodes $\Omega$, a single datapoint was chosen randomly and sampled with its given uncertainties, and the orbit scaled and rotated to pass through this point. The period was then calculated assuming a primary mass of $1.96\,\text{M}_\odot$. Each parameter set was accepted or rejected by comparing its likelihood to a number randomly drawn from a uniform probability distribution over the range (0,1). This process was repeated until 10000 orbits had been selected, from which we derived our posterior distributions. As in \citet{2017AJ....153..229B}, we increased the fraction of accepted orbits by normalising the likelihood function by the maximum likelihood found from an initial run of the algorithm. 
We find a 0.5$\degr$ offset in position angle between the NACO and SPHERE datasets taken in 2017, much larger than the expected change due to orbital motion and inconsistent with the 2016 and 2018 SPHERE measurements. Similar systematic offsets have been found in other datasets, and may originate from the different choice of astrometric reference fields and star catalogs used for the two instruments. To avoid any problems this may cause, we used only the SPHERE epochs to calculate the likelihood of potential orbits. In addition, we used only the astrometry measured in the IRDIS $H_2$ and $K_1$ filters, leading to 5 datapoints.

\begin{figure}
\includegraphics[width=0.49\textwidth]{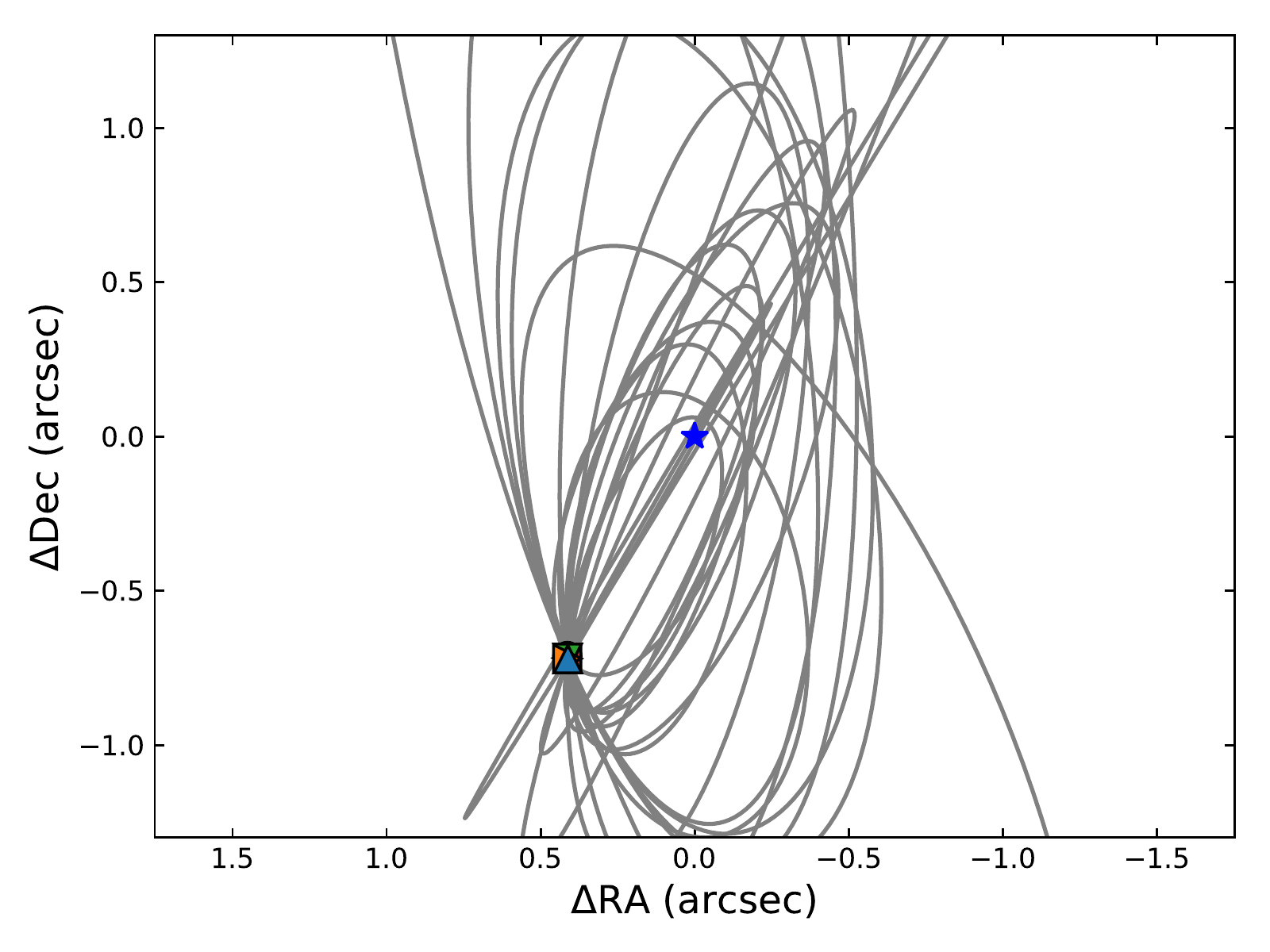}
\includegraphics[width=0.49\textwidth]{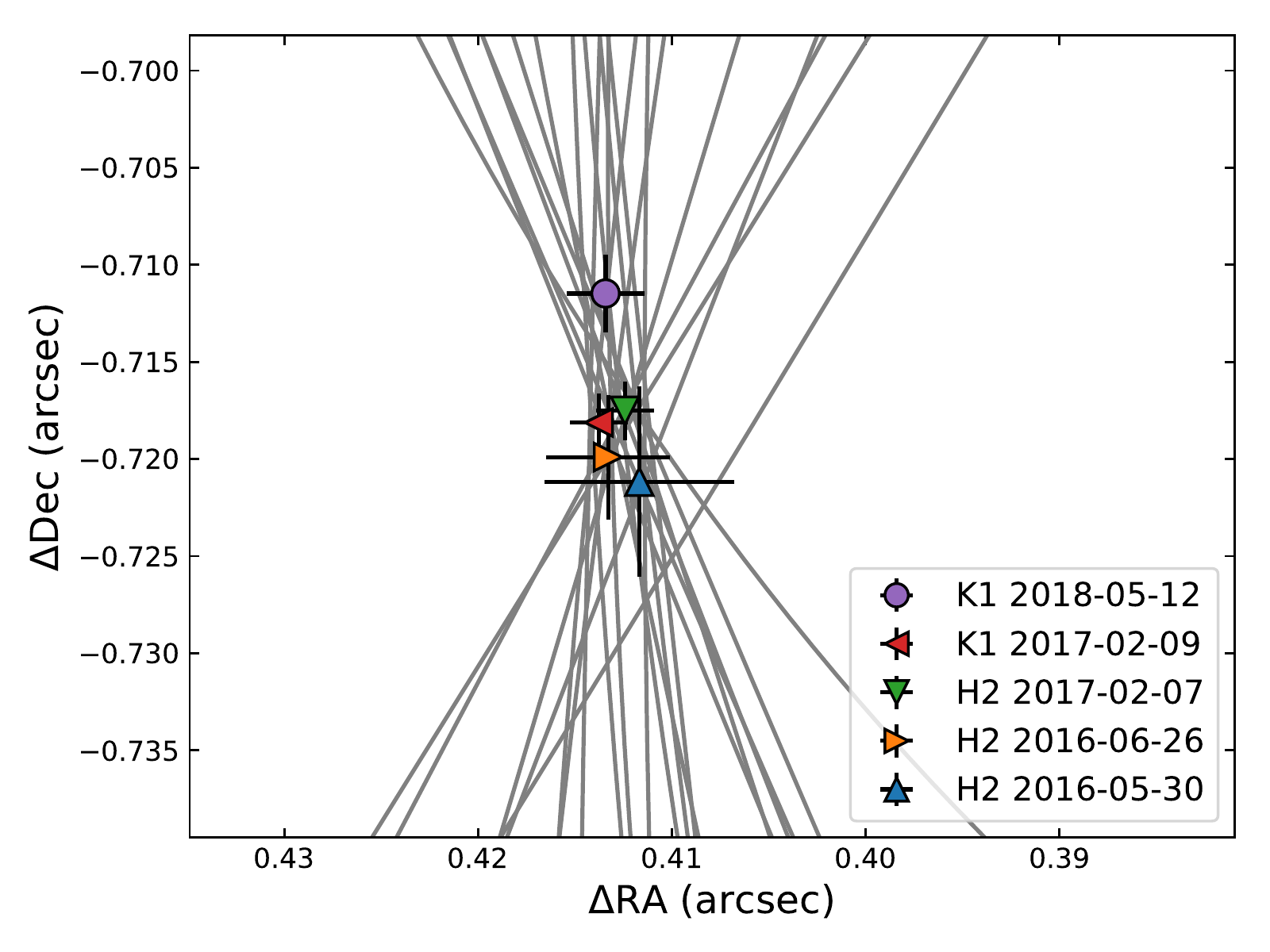}
\caption{The observed SPHERE-IRDIS astrometry of HIP\,65426\,b compared to 20 orbits drawn randomly from the 250 best-fitting orbits in the OFTI output sample. The position of the primary star is shown with a blue star. A close up of the region around the present position of HIP\,65426\,b is shown separately in the bottom plot.}\label{fig:orbit_plot}
\end{figure}

\begin{figure*}
\includegraphics[width=0.98\textwidth]{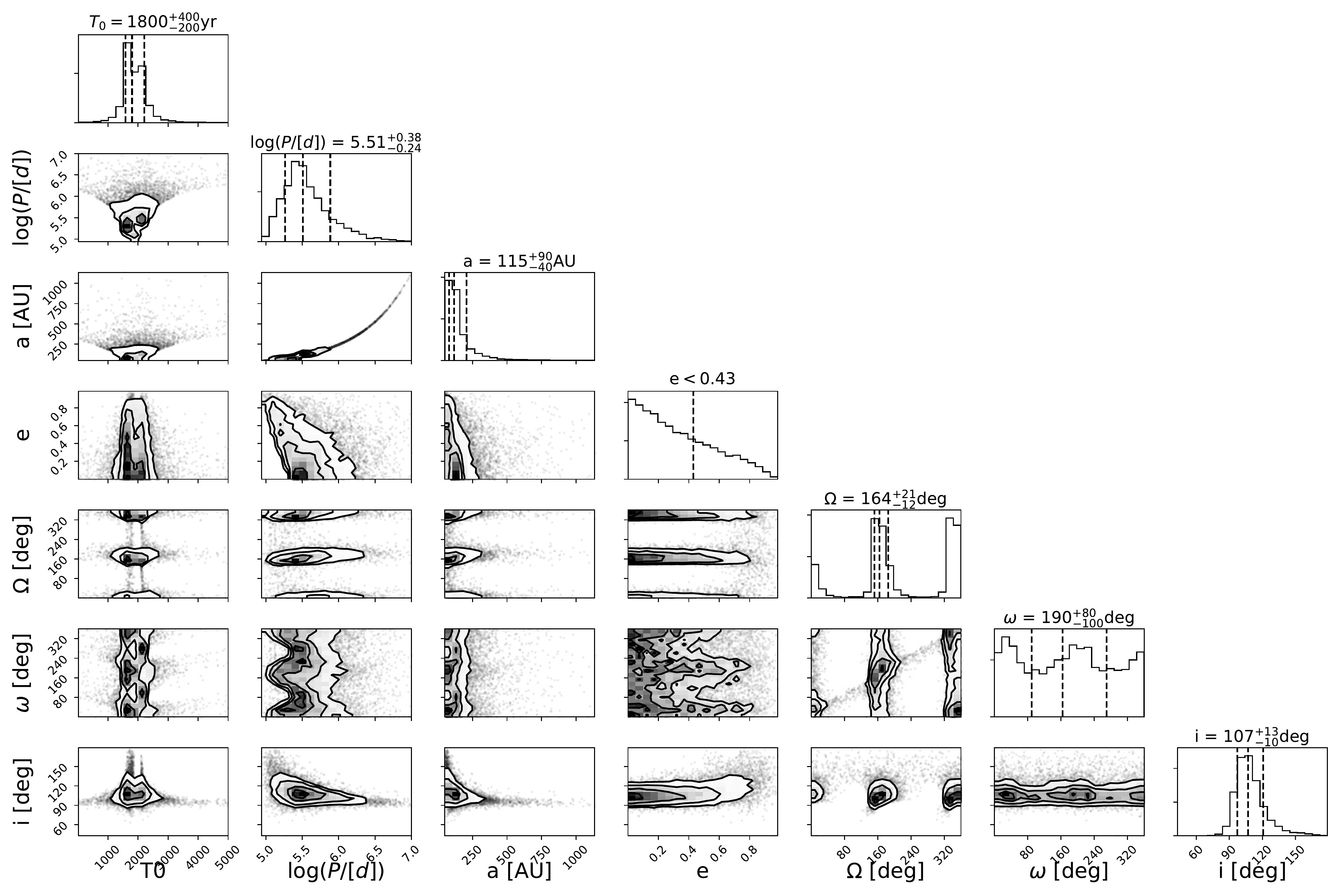}
\caption{Posterior distributions for the orbital parameters of HIP\,65426\,b around its host star. The dotted lines show the median value and 1$\sigma$ uncertainties for each parameter. For $\Omega$ and $\omega$, the value and uncertainties of one of the two solution families is plotted with a dotted line. For the eccentricity $e$, only the 1$\sigma$ limit is plotted.}\label{fig:orbit_posterior}
\end{figure*}

A sample of the resulting orbits are shown in Figure \ref{fig:orbit_plot}, while the full posterior distributions are shown in Figure \ref{fig:orbit_posterior}. From the posterior distributions, significant constraints can be placed on several of the orbital parameters, and the measured values are given in Table \ref{tab:orbit}. We find an orbital period of $\log(P[d]) = 5.51^{+0.37}_{-0.25}$ ($P=800^{+1200}_{-400}$\,yr), implying a semi-major axis of $a = 110^{+90}_{-30}$\,AU. Orbits close to edge-on are preferred, with a peak at $i=107^{+13}_{-10}$. While the $\sin i$ prior favours highly inclined orbits, the posterior distribution is significantly narrower than the distribution of generated orbits, showing that the inclination measurement is not dominated by the prior. Two families of solutions are found, with consistent parameters except for a 180 degree change in both $\Omega$ and $\omega$. This corresponds to our lack of knowledge as to whether HIP\,65426\,b is moving towards or away from us on its orbit. Due to the short time baseline compared to the orbital period and the lack of curvature in the observed motion, the eccentricity is essentially unconstrained, with the posterior distribution reflecting the linear prior used to draw sample orbits.  Since we used a linear prior, the peak occurs at e=0 and we instead express the eccentricity as a 1$\sigma$ upper limit ($e<0.43$). In addition, the orbital eccentricity shows significant correlations with several other parameters, with high eccentricity solutions preferring larger inclinations and shorter periods, for example. A different choice of prior would impact many of the orbital parameters.

\begin{table}
\caption{Measured orbital parameters of HIP\,65426\,b}
\label{tab:orbit}
\center
\begin{tabular}{ccc}
\hline
Parameter & Solution Family 1 & Solution Family 2\\
\hline
$\log(P[d])$ & $5.51^{+0.37}_{-0.25}$ & $5.50^{+0.38}_{-0.23}$ \\
$a$ [AU] & $120^{+90}_{-40}$  & $110^{+90}_{-30}$ \\
$T_0$ [year] & $1800^{+400}_{-200}$ & $1800^{+400}_{-200}$ \\
$e$ & $<0.43$ & $<0.43$ \\
$\Omega [\degr]$ & $164^{+21}_{-12}$ & $344^{+20}_{-12}$ \\
$\omega [\degr]$ & $190^{+80}_{-100}$  & $20^{+80}_{-90}$ \\
$i [\degr]$ & $107^{+13}_{-10}$  & $107^{+13}_{-10}$ \\
\hline
\end{tabular}
\end{table}

\subsection{Atmospheric models}

To estimate the physical parameters of HIP\,65426\,b we compared the observed spectrum to atmospheric models. We used the Markov chain Monte Carlo procedure described in \citet{2017A&A...603A..57S} to explore the likelihood distribution, interpolating the grid of spectral models at each step.

\begin{figure}
\includegraphics[width=0.5\textwidth]{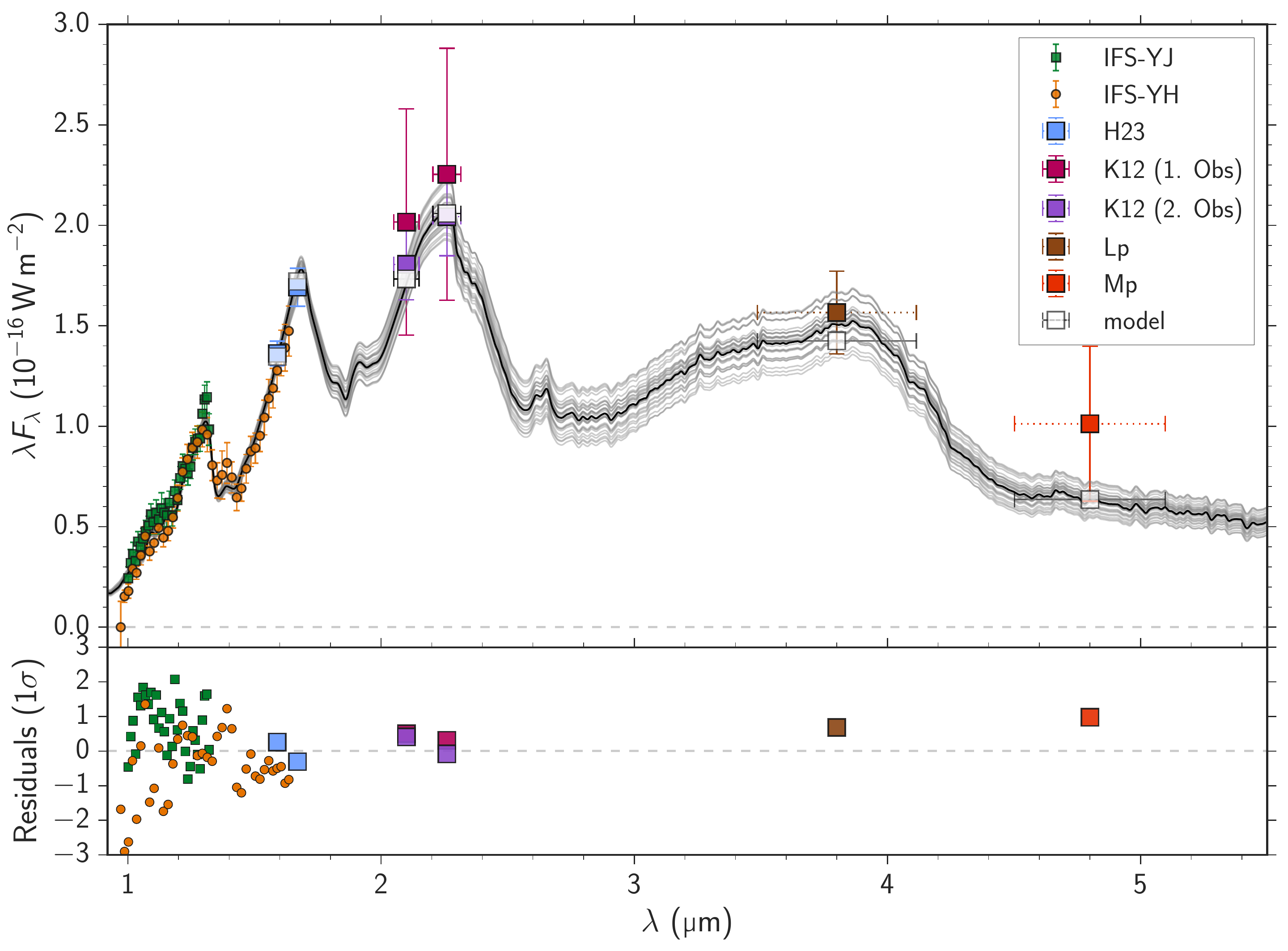}
\caption{The observed spectrum of HIP\,65426\,b (coloured squares) compared to the best-fit spectral model from the BT-Settl grid, with T$_{\text{eff}}$=1618\,K , $\log g=3.78$ and $R=1.17$R$_J$. The horizontal error bars show the width of the SPHERE and NACO filters, and the grey squares show the predicted flux calculated using the filter transmission curves. Several models drawn from the MCMC posterior distributions are also shown in grey.}\label{fig:spectrum}
\end{figure}

When comparing to the BT-Settl models \citep{2015A&A...577A..42B}, we found best-fit values similar to those of \citet{2017A&A...605L...9C}, with $T_{\text{eff}}=1618\pm7$\,K, $\log g=3.78^{+0.04}_{-0.03}$ dex and radius $R=1.17\pm0.04$\,$R_{\text{J}}$. This model is compared to the observed SED (spectral energy distribution) of HIP\,65426\,b in Figure \ref{fig:spectrum}.
Combining the temperature and radius measurements results in a luminosity estimate for HIP\,65426\,b of $\log \left(L/L_{\odot}\right)=-4.05\pm0.03$, while the radius and surface gravity produce a mass esimate of $M=3.32\pm0.4$\,$M_{\text{J}}$.
The listed uncertainties are optimistic, since they do not account for uncertainties in the modelling of these objects, and different models are likely to show systematic differences in the measured parameters that are larger than the given uncertainties. A number of free parameters are also not included in this analysis, including metallicity and cloud properties, that may impact the results.

We repeated the same procedure as above using the cloudy atmosphere petitCODE models, described in \citet{2017A&A...600A..10M}. We found that the best-fit solution converged to low temperatures ($\sim1200$\,K) with unphysically large radii ($>2$R$_{\text{J}}$), although a second likelihood peak at higher temperatures ($\sim1800$\,K) provided a similar fit to the data with different physical parameters. This set of models failed to simultaneously reproduce the shape of the J-band peak and the K-band flux, and so we concentrate on the BT-Settl results.

\subsection{Empirical Comparison}
Using the measured photometry, we placed HIP\,65426\,b on colour-colour and colour-magnitude diagrams. We compiled spectra of field dwarfs from \citet{2010ApJ...710.1627L}, the IRTF spectral library of M, L and T dwarfs \citep{2005ApJ...623.1115C,2009ApJS..185..289R} and the SpeX Prism Library \citep{2014ASInC..11....7B}. For targets where no parallax or distance was provided, we used the Database of Ultracool Parallaxes maintained by Trent Dupuy \citep{2012ApJS..201...19D,2013Sci...341.1492D} and the results of \citet{2017AJ....154..147D}, where available. We also included photometry from \citet{2012ApJS..201...19D}. For targets without $L'$ magnitudes, we converted {\it WISE W1} ($3.5$\,$\mu$m) absolute magnitudes to $L'$ ($3.8$\,$\mu$m) absolute magnitudes using a relation calculated from a linear fit to the targets with photometry in both bands. To compare the photometry of HIP\,65426\,b with those of other young targets, we included a range of directly imaged companions from the literature \citep{2004A&A...425L..29C,2013A&A...555A.107B,2015ApJ...811..103M,2016A&A...586L...8L,2016A&A...587A..56M,2016A&A...587A..58B,2016A&A...593A.119M,2017AJ....154...10R,2018arXiv180302725C}.

\begin{figure}
\includegraphics[width=0.5\textwidth]{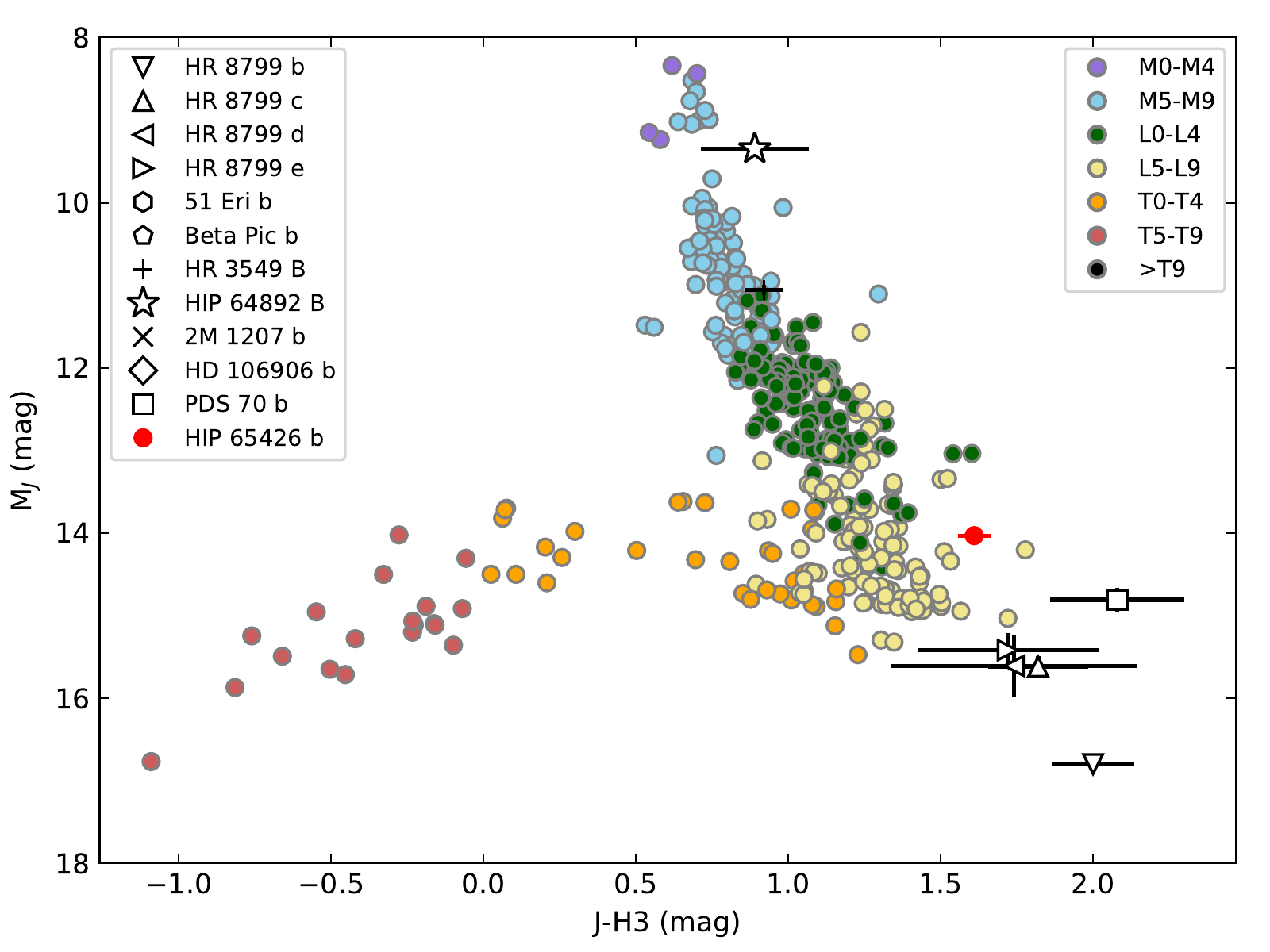}
\includegraphics[width=0.5\textwidth]{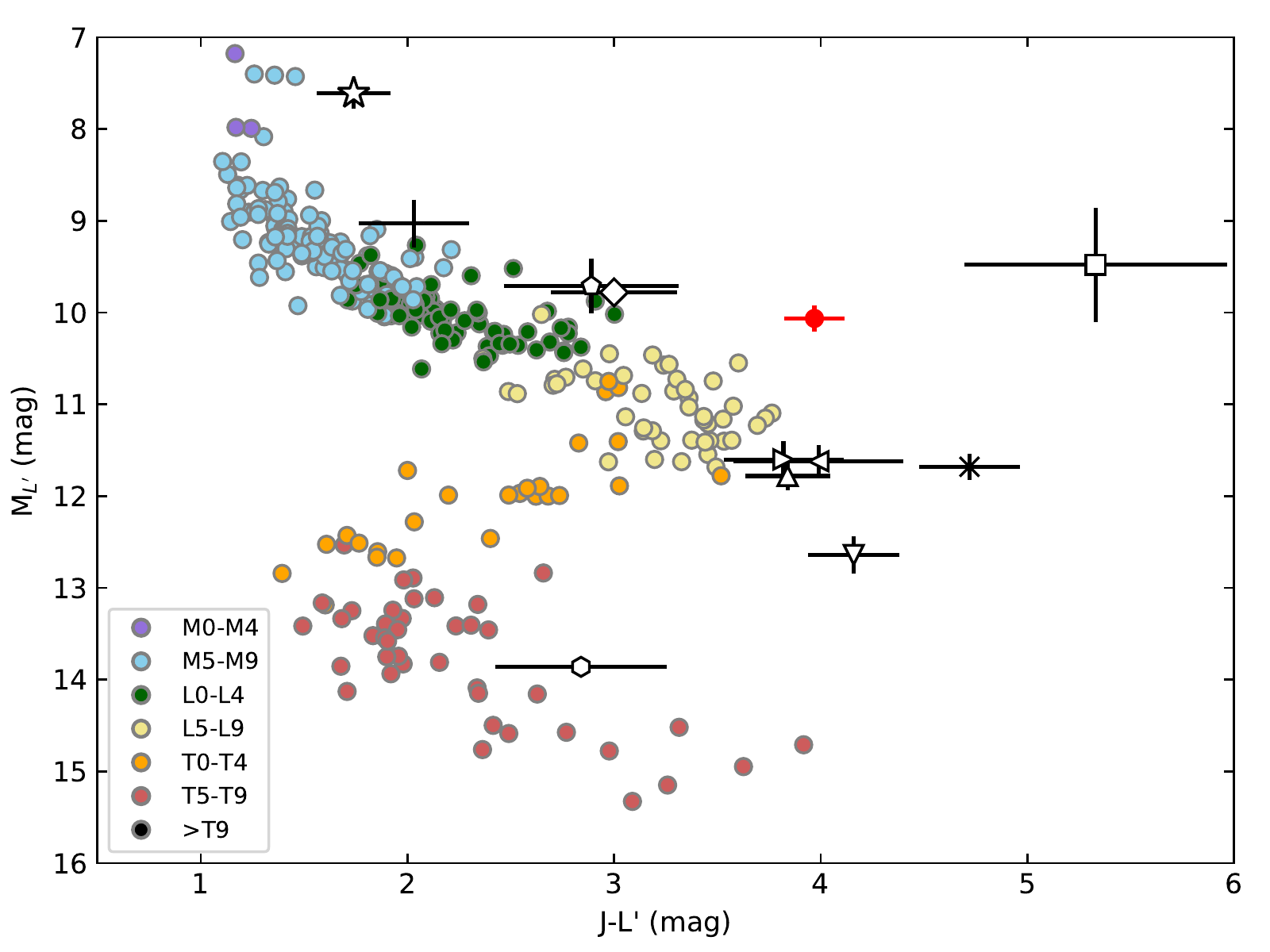}
\includegraphics[width=0.5\textwidth]{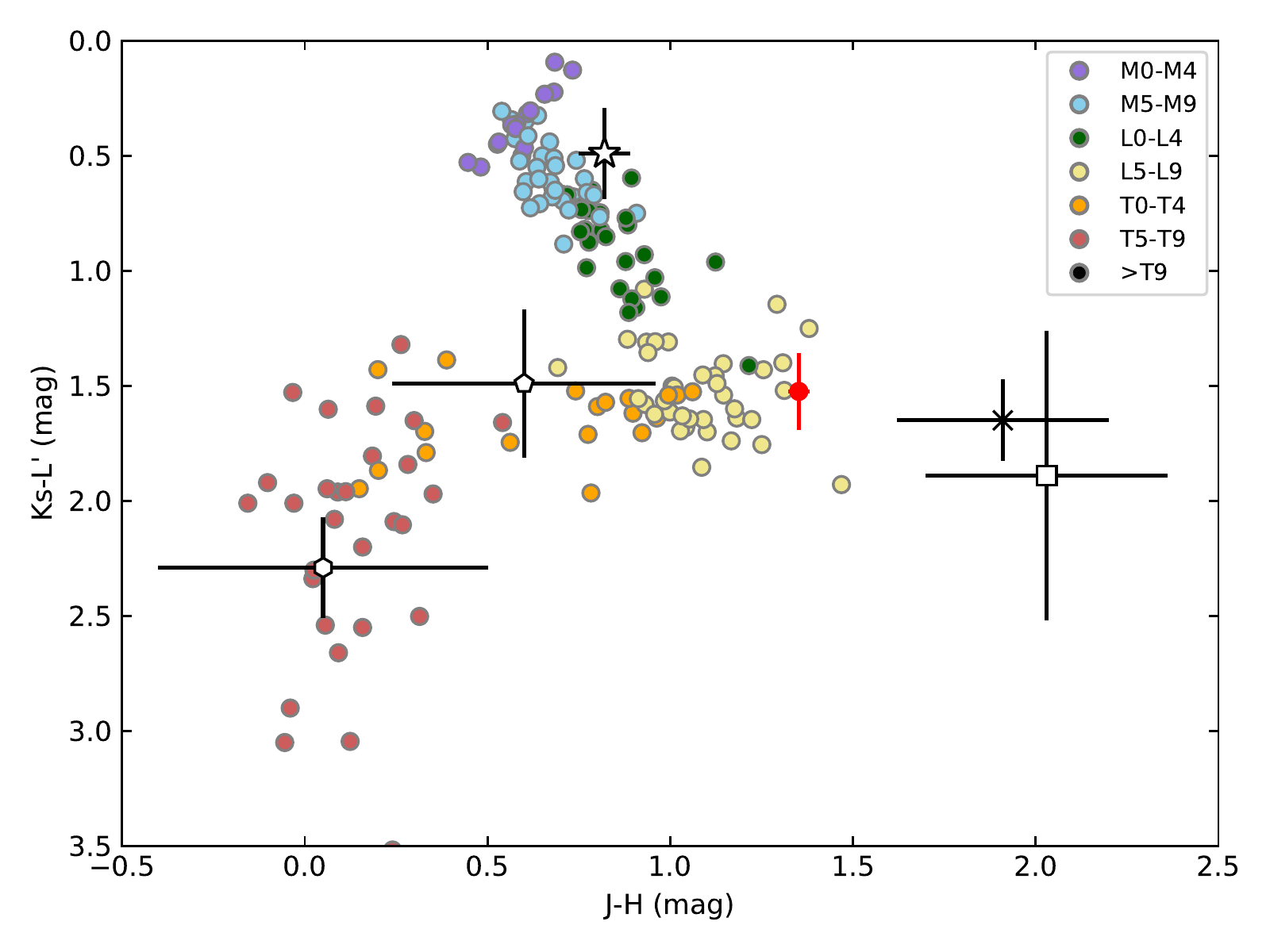}
\caption{Colour-magnitude and colour-colour diagrams showing the sequence of field dwarfs (coloured circles), a range of young, imaged brown dwarf and exoplanet companions (black symbols) and HIP\,65426\,b (red circle). HIP\,65426\,b falls near objects with mid-L spectral types, but is more luminous and with redder colours than the field sequence. This matches the trend seen for other young objects with similar spectral type.}\label{fig:color_mag_diags}
\end{figure}

From the colour-magnitude diagrams in Figure \ref{fig:color_mag_diags}, we can see that HIP\,65426\,b occupies a similar position to mid-late L-dwarfs. HIP\,65426\,b is redder and more luminous than the sequence of field dwarfs with similar spectral types. This property is typical for young field dwarfs \citep{2012ApJ...752...56F,2013AN....334...85L}, which contract and cool over time.

The position of HIP\,65426\,b on colour-magnitude diagrams is similar to that of PDS\,70\,b \citep{2018arXiv180611568K,2018arXiv180611567M}. At wavelengths longer than 1.4\,$\mu$m, their absolute magnitudes are similar. However, HIP\,65426\,b is significantly brighter in the J-band and displays systematically bluer colours than PDS\,70\,b, suggesting a higher temperature. Given that PDS\,70 is also significantly younger than HIP\,65426 \citep[$\sim5$\,Myr and $\sim14$\,Myr respectively;][]{2018arXiv180611568K,2017A&A...605L...9C}, the mass of HIP\,65426\,b is expected to be larger.

Compared to other planetary-mass companions, HIP\,65426\,b sits clearly between the early L-dwarfs (e.g. $\beta$ Pic b and HD 106906 b) and the L/T transition objects (e.g. HR 8799 c,d,e and 2M 1207 b). Together, these targets will be useful to study the atmospheric properties of young giant exoplanets across the L spectral sequence.

\subsection{Mass}\label{sec:mass}

To estimate the mass of HIP\,65426\,b, we compared the predictions of the DUSTY \citep{2000ApJ...542..464C} and COND \citep{2003A&A...402..701B} isochrones to the absolute magnitudes presented in Table \ref{tab:photometry}. We used a Monte Carlo procedure to include the uncertainty on the age of the HIP\,65426 system, drawing 10,000 samples from a Gaussian distribution consistent with the estimated age of $14\pm4$\,Myr. We interpolated the isochrones at each age and found the mass that gave the best match to the observed photometry, and took the median and standard deviation of these values as the mass and its uncertainty. With this method, the majority of samples fall below the minimum mass on the DUSTY model grid of $7.3$\,$M_{\text{J}}$, and so we estimate a 1$\sigma$ upper limit of $8.4$\,$M_{\text{J}}$. From the COND models we estimate a mass of $7.5\pm0.9$\,$M_{\text{J}}$.

We also used the luminosity estimated from the BT-SETTL model fit of $\log \left(L/L_{\odot}\right)=-4.05\pm0.03$, with the estimated age of $14\pm4$\,Myr. We find values of $8.2\pm1.1$\,$M_{\text{J}}$ and $8.3\pm0.9$\,$M_{\text{J}}$ for the DUSTY and COND grids respectively.

\begin{figure}
\includegraphics[width=0.5\textwidth]{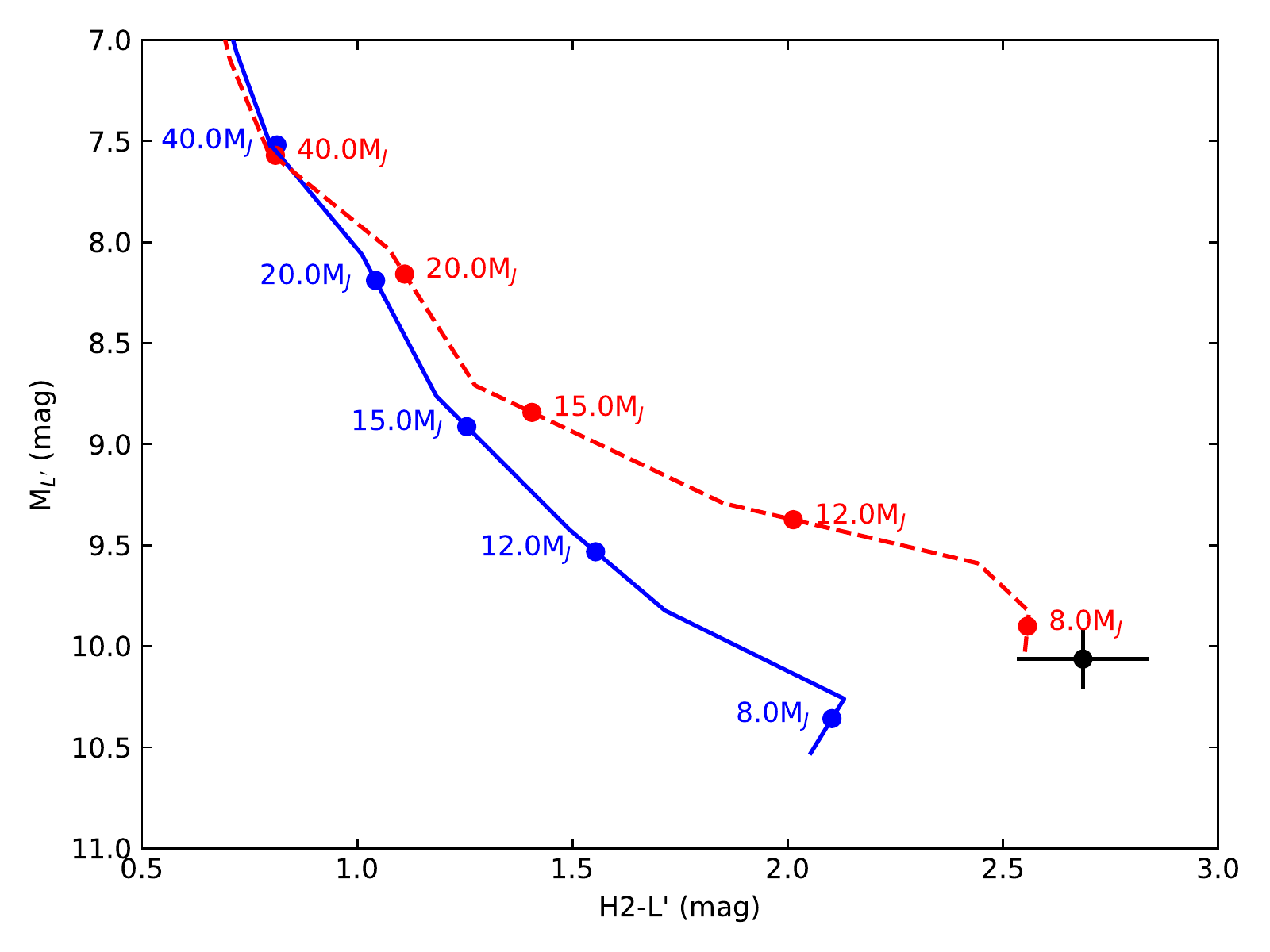}
\caption{Colour-magnitude diagram showing the COND (blue) and DUSTY (red) evolutionary models compared to HIP\,65426\,b. We find that the colours of HIP\,65426\,b match the DUSTY models more closely than the COND predictions.}\label{fig:color_mag_isochrone}
\end{figure}

We find that the colours of HIP\,65426\,b are systematically redder than those predicted by the COND isochrones. When placed on a range of colour-magnitude diagrams, such as the one in Figure \ref{fig:color_mag_isochrone}, the position of HIP\,65426\,b falls closer to the DUSTY tracks, suggesting the dusty nature of the companion.

An in-depth comparison of the photometry of HIP\,65426\,b with the BERN exoplanet cooling models by Marleau et al. (2018, submitted) resulted in mass estimates of $11.1^{+1.1}_{-2.2}$\,$M_{\text{J}}$ and $10.4^{+0.6}_{-2.6}$\,$M_{\text{J}}$ for their cold-start and hot-start simulations, suggesting a higher mass than the COND and DUSTY models.

\subsection{Limits on further companions}
\begin{figure}
\includegraphics[width=0.5\textwidth]{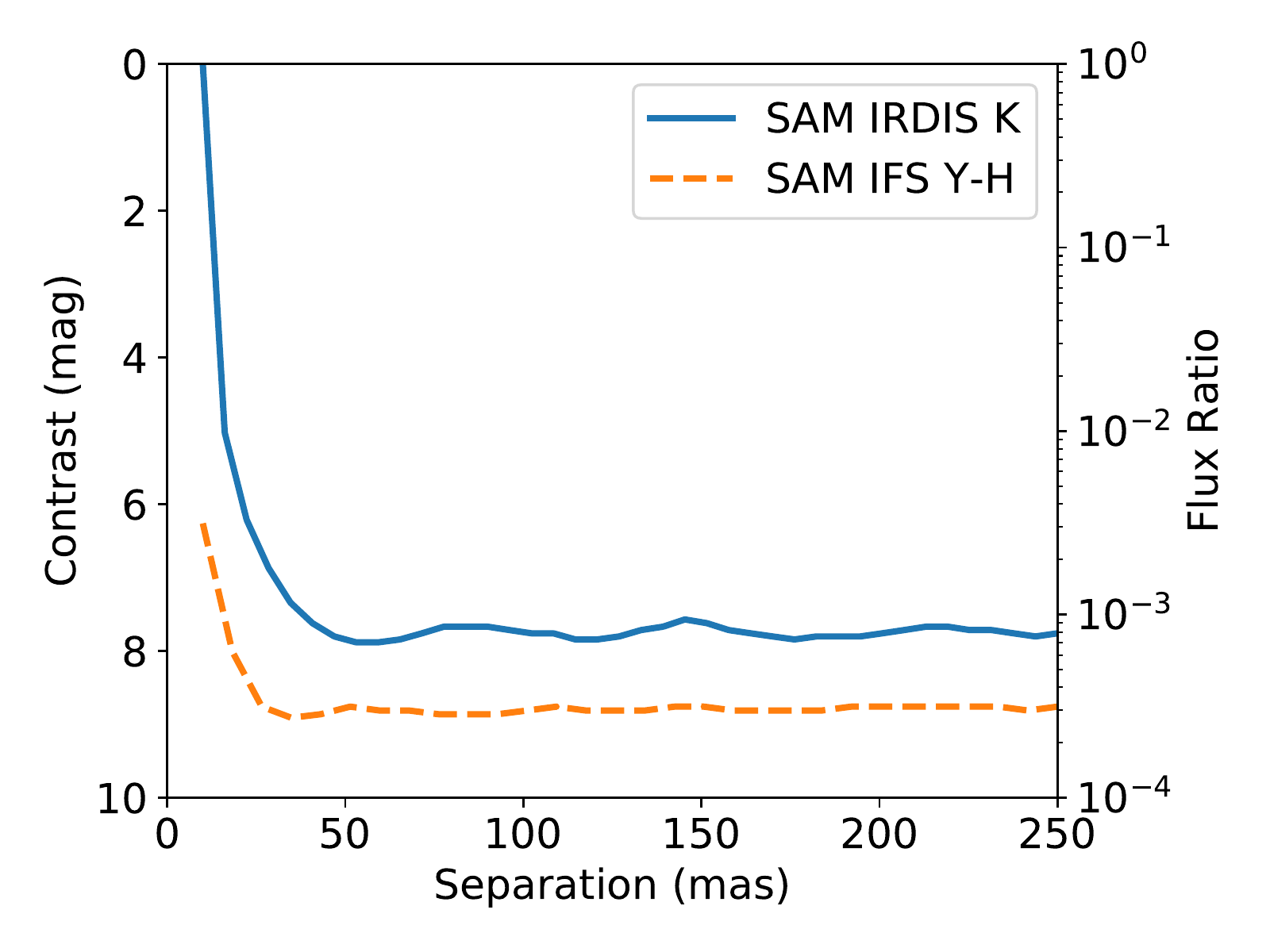}
\includegraphics[width=0.5\textwidth]{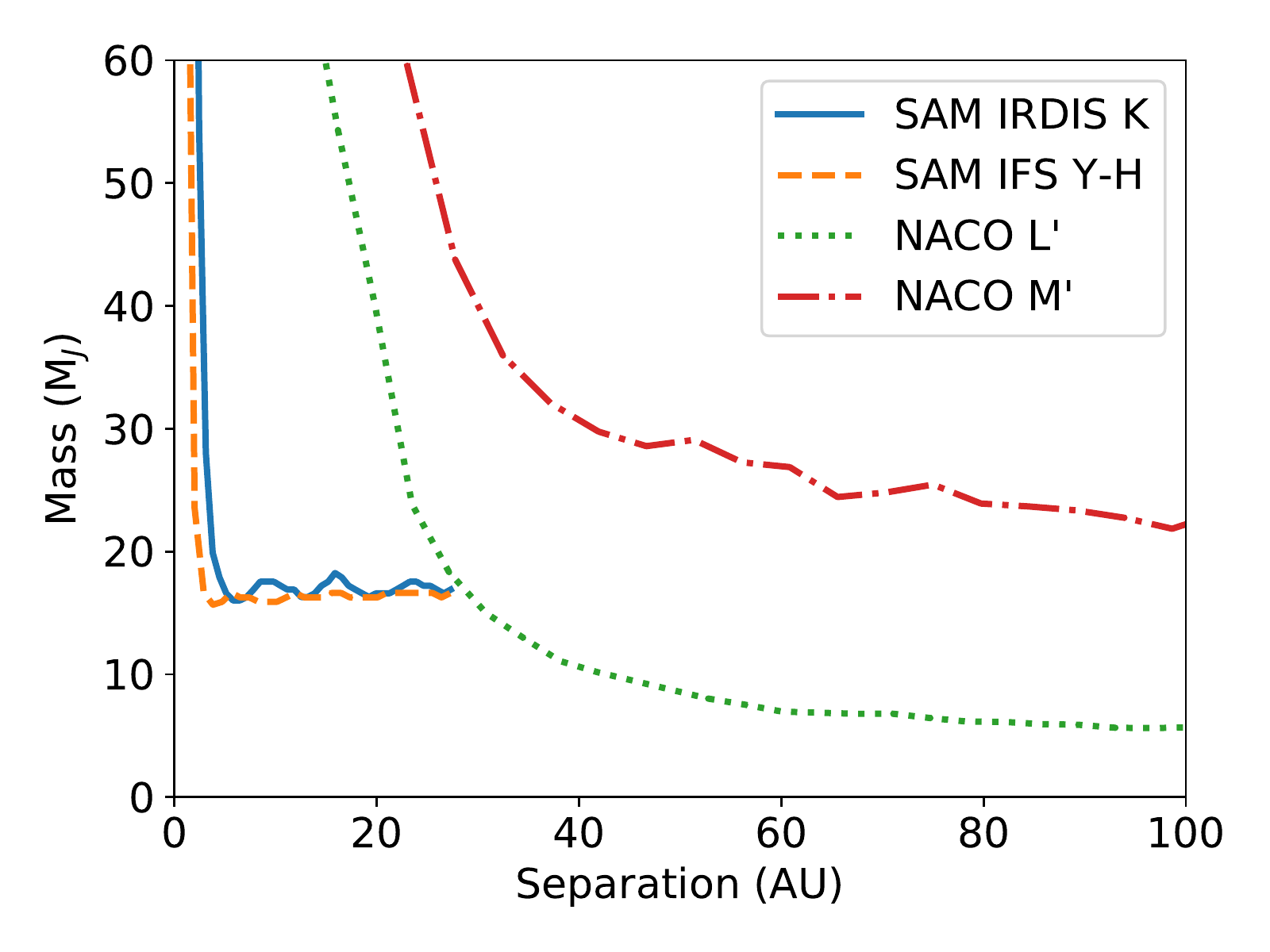}

\caption{Detection limits on further companions around HIP\,65426. Top: Detection limits from the SPHERE-SAM data in contrast and angular separation. Bottom: Mass limits for the SPHERE-SAM and NACO data expressed as projected physical separations. The limits in contrast were converted to masses using the DUSTY isochrones.}\label{fig:sam_detec_limits}
\end{figure}

We detect no additional companions in the NACO or SPHERE data. The SPHERE SAM detection limits probe a new parameter space not explored by \citet{2017A&A...605L...9C}. The 5$\sigma$ detection limits are displayed in Figure \ref{fig:sam_detec_limits}. Combining the detection limits with the DUSTY isochrones, we can rule out objects with masses larger than 20\,$M_{\text{J}}$ at separations of $>$2\,AU, and 16\,$M_{\text{J}}$ at $>$3\,AU, with a significance of $5\sigma$.

\section{Summary and Conclusions}
In this paper we have presented the first $L'$ and $M'$ images of the planetary companion HIP\,65426\,b, extending the coverage of its SED to the mid-infrared. The photometry extracted from these data are consistent with the conclusions of \citet{2017A&A...605L...9C}, confirming the nature of HIP\,65426\,b as a dusty, warm giant planet with a mid-L spectral type. Through comparison with the BT-Settl models, we find best-fit values of $T_{\text{eff}}=1618$\,K, surface gravity of $\log g=3.78$ and radius of $R=1.17$\,$R_{\text{J}}$. The observed SED is not fit well by the cloudy petitCODE models, with the grid unable to simultaneously reproduce the features seen at the $J$ and $K$ bands. Using the new thermal-infrared photometry, we update the estimated mass of HIP\,65426\,b, yielding a value of $8\pm1$\,$M_{\text{J}}$ from the DUSTY and COND isochrones. Among these two models, we find that its colours more closely match the predictions of the DUSTY models. This mass estimate is consistent with the results of Marleau et al. (2018, submitted) using the Bern exoplanet cooling models.

From the two-year time baseline between datasets, we have obtained the first indications of orbital motion of HIP\,65426\,b. We performed a preliminary orbital fit, with the present data preferring a close to edge-on orbit ($i=107^{+13}_{-10}$) with an orbital period of $P=800^{+1200}_{-400}$\,yr. Due to the large orbital period, a significantly longer time baseline will be needed for a full determination of its orbital parameters. 

By placing HIP\,65426\,b on colour-magnitude and colour-colour diagrams, we show that it fits in the sequence of other young giant planets. It displays redder colours and a higher luminosity than the field sequence, a trend typical of young, low gravity, and dusty objects. Amongst the other imaged planets, the recently discovered object PDS 70 b provides a close match to the observed spectrum of HIP\,65426\,b and may be a useful analog. These objects join a growing number of planetary-mass objects with L spectral types that have now been imaged. Given that low-surface gravity objects are expected to have cloud layers higher in their atmospheres \citep{2012ApJ...754..135M}, the L-T transition occurs at lower effective temperatures in young planets than field brown dwarfs. In this context, HIP\,65426\,b is an important object for understanding the atmospheric physics of young giant planets across the L spectral type where clouds are thought to play an important role, and for comparing their properties to that of the better-understood field brown dwarfs. 

As discussed in \citet{2017A&A...605L...9C}, the large separation of HIP\,65426\,b and its high mass are not consistent with in-situ formation via core accretion. However, the authors raised the possibility that it may have formed at a much smaller separation and experienced a scattering event with another companion, leading to its present wide orbit. This scenario was explored in detail by Marleau et al. (2018, submitted), who found that planet-planet scattering remains a viable pathway to produce objects like HIP\,65426\,b. Our SPHERE-SAM observations significantly narrow the parameter space for an additional massive object to occupy, ruling out the presence of $>16$\,$M_{\text{J}}$ brown dwarf companions at projected separations of $>$3\,AU. 
However, the current observations cannot discount the scenario of scattering caused by one or more additional giant planets on shorter-period orbits interior to that of HIP\,65426\,b.

\begin{acknowledgements}
This work has been carried out within the frame of the National Centre for Competence in Research ``PlanetS'' supported by the Swiss National Science Foundation (SNSF).\\
A.M. acknowledges the support of the DFG priority program SPP 1992 ``Exploring the Diversity of Extrasolar Planets'' (MU 4172/1-1). \\
J.\,O. and N.\,G. acknowledge financial support from the ICM (Iniciativa Cient\'ifica Milenio) via the N\'ucleo Milenio de Formaci\'on Planetaria grant. J.\,O. acknowledges financial support from the Universidad de Valpara\'iso, and from Fondecyt (grant 1180395). \\
A.Z. acknowledges support from the CONICYT + PAI/ Convocatoria nacional subvenci\'on a la instalaci\'on en la academia, convocatoria 2017 + Folio PAI77170087. N.G. acknowledges support from project CONICYT-PFCHA/Doctorado Nacional/2017 folio 21170650.\\
SPHERE is an instrument designed and built by a consortium consisting of IPAG (Grenoble, France), MPIA (Heidelberg, Germany), LAM (Marseille, France), LESIA (Paris, France), Laboratoire Lagrange (Nice, France), INAF - Osservatorio di Padova (Italy), Observatoire Astronomique de l'Université de Genève (Switzerland), ETH Zurich (Switzerland), NOVA (Netherlands), ONERA (France), and ASTRON (Netherlands) in collaboration with ESO. SPHERE was funded by ESO, with additional contributions from CNRS (France), MPIA (Germany), INAF (Italy), FINES (Switzerland), and NOVA (Netherlands). SPHERE also received funding from the European Commission Sixth and Seventh Framework Programmes as part of the Optical Infrared Coordination Network for Astronomy (OPTICON) under grant number RII3-Ct-2004-001566 for FP6 (2004-2008), grant number 226604 for FP7 (2009-2012), and grant number 312430 for FP7 (2013-2016). \\
This work has made use of the SPHERE Data Centre, jointly operated by OSUG/IPAG (Grenoble), PYTHEAS/LAM/CeSAM (Marseille), OCA/Lagrange (Nice), and Observatoire de Paris/LESIA (Paris) and supported by a grant from Labex OSUG@2020 (Investissements d’avenir – ANR10 LABX56).\\
This publication makes use of data products from the Two Micron All Sky Survey, which is a joint project of the University of Massachusetts and the Infrared Processing and Analysis Center/California Institute of Technology, funded by the National Aeronautics and Space Administration and the National Science Foundation.\\
This research has benefitted from the SpeX Prism Spectral Libraries, maintained by Adam Burgasser at \url {http://pono.ucsd.edu/~adam/browndwarfs/spexprism}. \\
This work has made use of data from the European Space Agency (ESA) mission
{\it Gaia} (\url{https://www.cosmos.esa.int/gaia}), processed by the {\it Gaia}
Data Processing and Analysis Consortium (DPAC,
\url{https://www.cosmos.esa.int/web/gaia/dpac/consortium}). Funding for the DPAC
has been provided by national institutions, in particular the institutions
participating in the {\it Gaia} Multilateral Agreement.\\
\end{acknowledgements}

\bibliographystyle{aa.bst}
\bibliography{big_bibliography.bib}

\begin{thebibliography}{71}
\expandafter\ifx\csname natexlab\endcsname\relax\def\natexlab#1{#1}\fi

\bibitem[{{Allard} {et~al.}(2012){Allard}, {Homeier}, \&
  {Freytag}}]{2012RSPTA.370.2765A}
{Allard}, F., {Homeier}, D., \& {Freytag}, B. 2012, Philosophical Transactions
  of the Royal Society of London Series A, 370, 2765

\bibitem[{{Amara} \& {Quanz}(2012)}]{2012MNRAS.427..948A}
{Amara}, A. \& {Quanz}, S.~P. 2012, \mnras, 427, 948

\bibitem[{{Baraffe} {et~al.}(2003){Baraffe}, {Chabrier}, {Barman}, {Allard}, \&
  {Hauschildt}}]{2003A&A...402..701B}
{Baraffe}, I., {Chabrier}, G., {Barman}, T.~S., {Allard}, F., \& {Hauschildt},
  P.~H. 2003, \aap, 402, 701

\bibitem[{{Baraffe} {et~al.}(2015){Baraffe}, {Homeier}, {Allard}, \&
  {Chabrier}}]{2015A&A...577A..42B}
{Baraffe}, I., {Homeier}, D., {Allard}, F., \& {Chabrier}, G. 2015, \aap, 577,
  A42

\bibitem[{{Beuzit} {et~al.}(2008){Beuzit}, {Feldt}, {Dohlen}, {Mouillet},
  {Puget}, {Wildi}, {Abe}, {Antichi}, {Baruffolo}, {Baudoz}, {Boccaletti},
  {Carbillet}, {Charton}, {Claudi}, {Downing}, {Fabron}, {Feautrier},
  {Fedrigo}, {Fusco}, {Gach}, {Gratton}, {Henning}, {Hubin}, {Joos}, {Kasper},
  {Langlois}, {Lenzen}, {Moutou}, {Pavlov}, {Petit}, {Pragt}, {Rabou}, {Rigal},
  {Roelfsema}, {Rousset}, {Saisse}, {Schmid}, {Stadler}, {Thalmann}, {Turatto},
  {Udry}, {Vakili}, \& {Waters}}]{2008SPIE.7014E..18B}
{Beuzit}, J.-L., {Feldt}, M., {Dohlen}, K., {et~al.} 2008, in \procspie, Vol.
  7014

\bibitem[{{Blunt} {et~al.}(2017){Blunt}, {Nielsen}, {De Rosa}, {Konopacky},
  {Ryan}, {Wang}, {Pueyo}, {Rameau}, {Marois}, {Marchis}, {Macintosh},
  {Graham}, {Duch{\^e}ne}, \& {Schneider}}]{2017AJ....153..229B}
{Blunt}, S., {Nielsen}, E.~L., {De Rosa}, R.~J., {et~al.} 2017, \aj, 153, 229

\bibitem[{{Boccaletti} {et~al.}(2015){Boccaletti}, {Lagage}, {Baudoz},
  {Beichman}, {Bouchet}, {Cavarroc}, {Dubreuil}, {Glasse}, {Glauser}, {Hines},
  {Lajoie}, {Lebreton}, {Perrin}, {Pueyo}, {Reess}, {Rieke}, {Ronayette},
  {Rouan}, {Soummer}, \& {Wright}}]{2015PASP..127..633B}
{Boccaletti}, A., {Lagage}, P.-O., {Baudoz}, P., {et~al.} 2015, \pasp, 127, 633

\bibitem[{{Bonnefoy} {et~al.}(2013){Bonnefoy}, {Boccaletti}, {Lagrange},
  {Allard}, {Mordasini}, {Beust}, {Chauvin}, {Girard}, {Homeier}, {Apai},
  {Lacour}, \& {Rouan}}]{2013A&A...555A.107B}
{Bonnefoy}, M., {Boccaletti}, A., {Lagrange}, A.-M., {et~al.} 2013, \aap, 555,
  A107

\bibitem[{{Bonnefoy} {et~al.}(2016){Bonnefoy}, {Zurlo}, {Baudino}, {Lucas},
  {Mesa}, {Maire}, {Vigan}, {Galicher}, {Homeier}, {Marocco}, {Gratton},
  {Chauvin}, {Allard}, {Desidera}, {Kasper}, {Moutou}, {Lagrange}, {Antichi},
  {Baruffolo}, {Baudrand}, {Beuzit}, {Boccaletti}, {Cantalloube}, {Carbillet},
  {Charton}, {Claudi}, {Costille}, {Dohlen}, {Dominik}, {Fantinel},
  {Feautrier}, {Feldt}, {Fusco}, {Gigan}, {Girard}, {Gluck}, {Gry}, {Henning},
  {Janson}, {Langlois}, {Madec}, {Magnard}, {Maurel}, {Mawet}, {Meyer},
  {Milli}, {Moeller-Nilsson}, {Mouillet}, {Pavlov}, {Perret}, {Pujet}, {Quanz},
  {Rochat}, {Rousset}, {Roux}, {Salasnich}, {Salter}, {Sauvage}, {Schmid},
  {Sevin}, {Soenke}, {Stadler}, {Turatto}, {Udry}, {Vakili}, {Wahhaj}, \&
  {Wildi}}]{2016A&A...587A..58B}
{Bonnefoy}, M., {Zurlo}, A., {Baudino}, J.~L., {et~al.} 2016, \aap, 587, A58

\bibitem[{{Burgasser}(2014)}]{2014ASInC..11....7B}
{Burgasser}, A.~J. 2014, in Astronomical Society of India Conference Series,
  Vol.~11, Astronomical Society of India Conference Series

\bibitem[{{Chabrier} {et~al.}(2000){Chabrier}, {Baraffe}, {Allard}, \&
  {Hauschildt}}]{2000ApJ...542..464C}
{Chabrier}, G., {Baraffe}, I., {Allard}, F., \& {Hauschildt}, P. 2000, \apj,
  542, 464

\bibitem[{{Chauvin} {et~al.}(2017{\natexlab{a}}){Chauvin}, {Desidera},
  {Lagrange}, {Vigan}, {Feldt}, {Gratton}, {Langlois}, {Cheetham}, {Bonnefoy},
  \& {Meyer}}]{2017sf2a.conf..331C}
{Chauvin}, G., {Desidera}, S., {Lagrange}, A.-M., {et~al.} 2017{\natexlab{a}},
  in SF2A-2017: Proceedings of the Annual meeting of the French Society of
  Astronomy and Astrophysics, ed. C.~{Reyl{\'e}}, P.~{Di Matteo}, F.~{Herpin},
  E.~{Lagadec}, A.~{Lan{\c c}on}, Z.~{Meliani}, \& F.~{Royer}, 331--335

\bibitem[{{Chauvin} {et~al.}(2017{\natexlab{b}}){Chauvin}, {Desidera},
  {Lagrange}, {Vigan}, {Gratton}, {Langlois}, {Bonnefoy}, {Beuzit}, {Feldt},
  {Mouillet}, {Meyer}, {Cheetham}, {Biller}, {Boccaletti}, {D'Orazi},
  {Galicher}, {Hagelberg}, {Maire}, {Mesa}, {Olofsson}, {Samland}, {Schmidt},
  {Sissa}, {Bonavita}, {Charnay}, {Cudel}, {Daemgen}, {Delorme},
  {Janin-Potiron}, {Janson}, {Keppler}, {Le Coroller}, {Ligi}, {Marleau},
  {Messina}, {Molli{\`e}re}, {Mordasini}, {M{\"u}ller}, {Peretti}, {Perrot},
  {Rodet}, {Rouan}, {Zurlo}, {Dominik}, {Henning}, {Menard}, {Schmid},
  {Turatto}, {Udry}, {Vakili}, {Abe}, {Antichi}, {Baruffolo}, {Baudoz},
  {Baudrand}, {Blanchard}, {Bazzon}, {Buey}, {Carbillet}, {Carle}, {Charton},
  {Cascone}, {Claudi}, {Costille}, {Deboulbe}, {De Caprio}, {Dohlen},
  {Fantinel}, {Feautrier}, {Fusco}, {Gigan}, {Giro}, {Gisler}, {Gluck},
  {Hubin}, {Hugot}, {Jaquet}, {Kasper}, {Madec}, {Magnard}, {Martinez},
  {Maurel}, {Le Mignant}, {M{\"o}ller-Nilsson}, {Llored}, {Moulin},
  {Orign{\'e}}, {Pavlov}, {Perret}, {Petit}, {Pragt}, {Puget}, {Rabou},
  {Ramos}, {Rigal}, {Rochat}, {Roelfsema}, {Rousset}, {Roux}, {Salasnich},
  {Sauvage}, {Sevin}, {Soenke}, {Stadler}, {Suarez}, {Weber}, {Wildi},
  {Antoniucci}, {Augereau}, {Baudino}, {Brandner}, {Engler}, {Girard}, {Gry},
  {Kral}, {Kopytova}, {Lagadec}, {Milli}, {Moutou}, {Schlieder},
  {Szul{\'a}gyi}, {Thalmann}, \& {Wahhaj}}]{2017A&A...605L...9C}
{Chauvin}, G., {Desidera}, S., {Lagrange}, A.-M., {et~al.} 2017{\natexlab{b}},
  \aap, 605, L9

\bibitem[{{Chauvin} {et~al.}(2012){Chauvin}, {Lagrange}, {Beust}, {Bonnefoy},
  {Boccaletti}, {Apai}, {Allard}, {Ehrenreich}, {Girard}, {Mouillet}, \&
  {Rouan}}]{2012A&A...542A..41C}
{Chauvin}, G., {Lagrange}, A.-M., {Beust}, H., {et~al.} 2012, \aap, 542, A41

\bibitem[{{Chauvin} {et~al.}(2004){Chauvin}, {Lagrange}, {Dumas}, {Zuckerman},
  {Mouillet}, {Song}, {Beuzit}, \& {Lowrance}}]{2004A&A...425L..29C}
{Chauvin}, G., {Lagrange}, A.-M., {Dumas}, C., {et~al.} 2004, \aap, 425, L29

\bibitem[{{Cheetham} {et~al.}(2018){Cheetham}, {Bonnefoy}, {Desidera},
  {Langlois}, {Vigan}, {Schmidt}, {Olofsson}, {Chauvin}, {Klahr}, {Gratton},
  {D'Orazi}, {Henning}, {Janson}, {Biller}, {Peretti}, {Hagelberg},
  {S{\'e}gransan}, {Udry}, {Mesa}, {Sissa}, {Kral}, {Schlieder}, {Maire},
  {Mordasini}, {Menard}, {Zurlo}, {Beuzit}, {Feldt}, {Mouillet}, {Meyer},
  {Lagrange}, {Boccaletti}, {Keppler}, {Kopytova}, {Ligi}, {Rouan}, {Le
  Coroller}, {Dominik}, {Lagadec}, {Turatto}, {Abe}, {Antichi}, {Baruffolo},
  {Baudoz}, {Blanchard}, {Buey}, {Carbillet}, {Carle}, {Cascone}, {Claudi},
  {Costille}, {Delboulb{\'e}}, {De Caprio}, {Dohlen}, {Fantinel}, {Feautrier},
  {Fusco}, {Giro}, {Gluck}, {Hubin}, {Hugot}, {Jaquet}, {Kasper}, {Llored},
  {Madec}, {Magnard}, {Martinez}, {Maurel}, {Le Mignant}, {M{\"o}ller-Nilsson},
  {Moulin}, {Orign{\'e}}, {Pavlov}, {Perret}, {Petit}, {Pragt}, {Puget},
  {Rabou}, {Ramos}, {Rigal}, {Rochat}, {Roelfsema}, {Rousset}, {Roux},
  {Salasnich}, {Sauvage}, {Sevin}, {Soenke}, {Stadler}, {Suarez}, {Weber}, \&
  {Wildi}}]{2018arXiv180302725C}
{Cheetham}, A., {Bonnefoy}, M., {Desidera}, S., {et~al.} 2018, ArXiv e-prints
  [\eprint[arXiv]{1803.02725}]

\bibitem[{{Cheetham} {et~al.}(2016){Cheetham}, {Girard}, {Lacour}, {Schworer},
  {Haubois}, \& {Beuzit}}]{2016SPIE.9907E..2TC}
{Cheetham}, A.~C., {Girard}, J., {Lacour}, S., {et~al.} 2016, in \procspie,
  Vol. 9907, Optical and Infrared Interferometry and Imaging V, 99072T

\bibitem[{{Claudi} {et~al.}(2008){Claudi}, {Turatto}, {Gratton}, {Antichi},
  {Bonavita}, {Bruno}, {Cascone}, {De Caprio}, {Desidera}, {Giro}, {Mesa},
  {Scuderi}, {Dohlen}, {Beuzit}, \& {Puget}}]{2008SPIE.7014E..3EC}
{Claudi}, R.~U., {Turatto}, M., {Gratton}, R.~G., {et~al.} 2008, in \procspie,
  Vol. 7014, Ground-based and Airborne Instrumentation for Astronomy II, 70143E

\bibitem[{{Currie} {et~al.}(2014){Currie}, {Burrows}, {Girard}, {Cloutier},
  {Fukagawa}, {Sorahana}, {Kuchner}, {Kenyon}, {Madhusudhan}, {Itoh},
  {Jayawardhana}, {Matsumura}, \& {Pyo}}]{2014ApJ...795..133C}
{Currie}, T., {Burrows}, A., {Girard}, J.~H., {et~al.} 2014, \apj, 795, 133

\bibitem[{{Cushing} {et~al.}(2005){Cushing}, {Rayner}, \&
  {Vacca}}]{2005ApJ...623.1115C}
{Cushing}, M.~C., {Rayner}, J.~T., \& {Vacca}, W.~D. 2005, \apj, 623, 1115

\bibitem[{{Dahn} {et~al.}(2017){Dahn}, {Harris}, {Subasavage}, {Ables},
  {Canzian}, {Guetter}, {Harris}, {Henden}, {Leggett}, {Levine}, {Luginbuhl},
  {Monet}, {Monet}, {Munn}, {Pier}, {Stone}, {Vrba}, {Walker}, \&
  {Tilleman}}]{2017AJ....154..147D}
{Dahn}, C.~C., {Harris}, H.~C., {Subasavage}, J.~P., {et~al.} 2017, \aj, 154,
  147

\bibitem[{{Dohlen} {et~al.}(2008){Dohlen}, {Langlois}, {Saisse}, {Hill},
  {Origne}, {Jacquet}, {Fabron}, {Blanc}, {Llored}, {Carle}, {Moutou}, {Vigan},
  {Boccaletti}, {Carbillet}, {Mouillet}, \& {Beuzit}}]{2008SPIE.7014E..3LD}
{Dohlen}, K., {Langlois}, M., {Saisse}, M., {et~al.} 2008, in \procspie, Vol.
  7014, Ground-based and Airborne Instrumentation for Astronomy II, 70143L

\bibitem[{{Dupuy} \& {Kraus}(2013)}]{2013Sci...341.1492D}
{Dupuy}, T.~J. \& {Kraus}, A.~L. 2013, Science, 341, 1492

\bibitem[{{Dupuy} \& {Liu}(2012)}]{2012ApJS..201...19D}
{Dupuy}, T.~J. \& {Liu}, M.~C. 2012, \apjs, 201, 19

\bibitem[{{Faherty} {et~al.}(2012){Faherty}, {Burgasser}, {Walter}, {Van der
  Bliek}, {Shara}, {Cruz}, {West}, {Vrba}, \&
  {Anglada-Escud{\'e}}}]{2012ApJ...752...56F}
{Faherty}, J.~K., {Burgasser}, A.~J., {Walter}, F.~M., {et~al.} 2012, \apj,
  752, 56

\bibitem[{{Gaia Collaboration} {et~al.}(2018){Gaia Collaboration}, {Brown},
  {Vallenari}, {Prusti}, {de Bruijne}, {Babusiaux}, \&
  {Bailer-Jones}}]{2018arXiv180409365G}
{Gaia Collaboration}, {Brown}, A.~G.~A., {Vallenari}, A., {et~al.} 2018, ArXiv
  e-prints [\eprint[arXiv]{1804.09365}]

\bibitem[{{Gaia Collaboration} {et~al.}(2016){Gaia Collaboration}, {Prusti},
  {de Bruijne}, {Brown}, {Vallenari}, {Babusiaux}, {Bailer-Jones}, {Bastian},
  {Biermann}, {Evans}, \& et~al.}]{2016A&A...595A...1G}
{Gaia Collaboration}, {Prusti}, T., {de Bruijne}, J.~H.~J., {et~al.} 2016,
  \aap, 595, A1

\bibitem[{{Galicher} {et~al.}(2018){Galicher}, {Boccaletti}, {Mesa}, {Delorme},
  {Gratton}, {Langlois}, {Lagrange}, {Maire}, {Le Coroller}, {Chauvin},
  {Biller}, {Cantalloube}, {Janson}, {Lagadec}, {Meunier}, {Vigan},
  {Hagelberg}, {Bonnefoy}, {Zurlo}, {Rocha}, {Maurel}, {Jaquet}, {Buey}, \&
  {Weber}}]{2018A&A...615A..92G}
{Galicher}, R., {Boccaletti}, A., {Mesa}, D., {et~al.} 2018, \aap, 615, A92

\bibitem[{{Galicher} {et~al.}(2011){Galicher}, {Marois}, {Macintosh}, {Barman},
  \& {Konopacky}}]{2011ApJ...739L..41G}
{Galicher}, R., {Marois}, C., {Macintosh}, B., {Barman}, T., \& {Konopacky}, Q.
  2011, \apjl, 739, L41

\bibitem[{{Hagelberg} {et~al.}(2016){Hagelberg}, {S{\'e}gransan}, {Udry}, \&
  {Wildi}}]{2016MNRAS.455.2178H}
{Hagelberg}, J., {S{\'e}gransan}, D., {Udry}, S., \& {Wildi}, F. 2016, \mnras,
  455, 2178

\bibitem[{{Hinkley} {et~al.}(2017){Hinkley}, {Skemer}, {Biller}, {Baraffe},
  {Bonnefoy}, {Bowler}, {Carter}, {Chen}, {Choquet}, {Currie}, {Danielski},
  {Fortney}, {Grady}, {Greenbaum}, {Hines}, {Janson}, {Kalas}, {Kennedy},
  {Kraus}, {Lagrange}, {Liu}, {Marley}, {Marois}, {Matthews}, {Mawet},
  {Metchev}, {Meyer}, {Millar-Blanchaer}, {Perrin}, {Pueyo}, {Quanz}, {Rameau},
  {Rodigas}, {Sallum}, {Sargent}, {Schlieder}, {Schneider}, {Stapelfeldt},
  {Tremblin}, {Vigan}, \& {Ygouf}}]{2017jwst.prop.1386H}
{Hinkley}, S., {Skemer}, A., {Biller}, B., {et~al.} 2017, {High Contrast
  Imaging of Exoplanets and Exoplanetary Systems with JWST}, JWST Proposal ID
  1386. Cycle 0 Early Release Science

\bibitem[{{H{\o}g} {et~al.}(2000){H{\o}g}, {Fabricius}, {Makarov}, {Urban},
  {Corbin}, {Wycoff}, {Bastian}, {Schwekendiek}, \&
  {Wicenec}}]{2000AA...355L..27H}
{H{\o}g}, E., {Fabricius}, C., {Makarov}, V.~V., {et~al.} 2000, \aap, 355, L27

\bibitem[{{Hunziker} {et~al.}(2017){Hunziker}, {Quanz}, {Amara}, \&
  {Meyer}}]{2017arXiv170610069H}
{Hunziker}, S., {Quanz}, S.~P., {Amara}, A., \& {Meyer}, M.~R. 2017, ArXiv
  e-prints [\eprint[arXiv]{1706.10069}]

\bibitem[{{Jennison}(1958)}]{1958MNRAS.118..276J}
{Jennison}, R.~C. 1958, \mnras, 118, 276

\bibitem[{{Keppler} {et~al.}(2018){Keppler}, {Benisty}, {M{\"u}ller},
  {Henning}, {van Boekel}, {Cantalloube}, {Ginski}, {van Holstein}, {Maire},
  {Pohl}, {Samland}, {Avenhaus}, {Baudino}, {Boccaletti}, {de Boer},
  {Bonnefoy}, {Chauvin}, {Desidera}, {Langlois}, {Lazzoni}, {Marleau},
  {Mordasini}, {Pawellek}, {Stolker}, {Vigan}, {Zurlo}, {Birnstiel},
  {Brandner}, {Feldt}, {Flock}, {Girard}, {Gratton}, {Hagelberg}, {Isella},
  {Janson}, {Juhasz}, {Kemmer}, {Kral}, {Lagrange}, {Launhardt}, {Matter},
  {M{\'e}nard}, {Milli}, {Molli{\`e}re}, {Olofsson}, {Perez}, {Pinilla},
  {Pinte}, {Quanz}, {Schmidt}, {Udry}, {Wahhaj}, {Williams}, {Buenzli},
  {Cudel}, {Dominik}, {Galicher}, {Kasper}, {Lannier}, {Mesa}, {Mouillet},
  {Peretti}, {Perrot}, {Salter}, {Sissa}, {Wildi}, {Abe}, {Antichi},
  {Augereau}, {Baruffolo}, {Baudoz}, {Bazzon}, {Beuzit}, {Blanchard}, {Brems},
  {Buey}, {De Caprio}, {Carbillet}, {Carle}, {Cascone}, {Cheetham}, {Claudi},
  {Costille}, {Delboulb{\'e}}, {Dohlen}, {Fantinel}, {Feautrier}, {Fusco},
  {Giro}, {Gisler}, {Gluck}, {Gry}, {Hubin}, {Hugot}, {Jaquet}, {Le Mignant},
  {Llored}, {Madec}, {Magnard}, {Martinez}, {Maurel}, {Meyer},
  {Moeller-Nilsson}, {Moulin}, {Mugnier}, {Origne}, {Pavlov}, {Perret},
  {Petit}, {Pragt}, {Puget}, {Rabou}, {Ramos}, {Rigal}, {Rochat}, {Roelfsema},
  {Rousset}, {Roux}, {Salasnich}, {Sauvage}, {Sevin}, {Soenke}, {Stadler},
  {Suarez}, {Turatto}, \& {Weber}}]{2018arXiv180611568K}
{Keppler}, M., {Benisty}, M., {M{\"u}ller}, A., {et~al.} 2018, ArXiv e-prints
  [\eprint[arXiv]{1806.11568}]

\bibitem[{{Kraus} {et~al.}(2008){Kraus}, {Ireland}, {Martinache}, \&
  {Lloyd}}]{kraus08}
{Kraus}, A.~L., {Ireland}, M.~J., {Martinache}, F., \& {Lloyd}, J.~P. 2008,
  \apj, 679, 762

\bibitem[{{Krist} {et~al.}(2007){Krist}, {Beichman}, {Trauger}, {Rieke},
  {Somerstein}, {Green}, {Horner}, {Stansberry}, {Shi}, {Meyer}, {Stapelfeldt},
  \& {Roellig}}]{2007SPIE.6693E..0HK}
{Krist}, J.~E., {Beichman}, C.~A., {Trauger}, J.~T., {et~al.} 2007, in
  \procspie, Vol. 6693, Techniques and Instrumentation for Detection of
  Exoplanets III, 66930H

\bibitem[{{Lagrange} {et~al.}(2010){Lagrange}, {Bonnefoy}, {Chauvin}, {Apai},
  {Ehrenreich}, {Boccaletti}, {Gratadour}, {Rouan}, {Mouillet}, {Lacour}, \&
  {Kasper}}]{2010Sci...329...57L}
{Lagrange}, A.-M., {Bonnefoy}, M., {Chauvin}, G., {et~al.} 2010, Science, 329,
  57

\bibitem[{{Lagrange} {et~al.}(2016){Lagrange}, {Langlois}, {Gratton}, {Maire},
  {Milli}, {Olofsson}, {Vigan}, {Bailey}, {Mesa}, {Chauvin}, {Boccaletti},
  {Galicher}, {Girard}, {Bonnefoy}, {Samland}, {Menard}, {Henning},
  {Kenworthy}, {Thalmann}, {Beust}, {Beuzit}, {Brandner}, {Buenzli},
  {Cheetham}, {Janson}, {le Coroller}, {Lannier}, {Mouillet}, {Peretti},
  {Perrot}, {Salter}, {Sissa}, {Wahhaj}, {Abe}, {Desidera}, {Feldt}, {Madec},
  {Perret}, {Petit}, {Rabou}, {Soenke}, \& {Weber}}]{2016A&A...586L...8L}
{Lagrange}, A.-M., {Langlois}, M., {Gratton}, R., {et~al.} 2016, \aap, 586, L8

\bibitem[{Larkin {et~al.}(1997)Larkin, Oldfield, \& Klemm}]{larkin1997fast}
Larkin, K.~G., Oldfield, M.~A., \& Klemm, H. 1997, Optics communications, 139,
  99

\bibitem[{{Leggett} {et~al.}(2010){Leggett}, {Burningham}, {Saumon}, {Marley},
  {Warren}, {Smart}, {Jones}, {Lucas}, {Pinfield}, \&
  {Tamura}}]{2010ApJ...710.1627L}
{Leggett}, S.~K., {Burningham}, B., {Saumon}, D., {et~al.} 2010, \apj, 710,
  1627

\bibitem[{{Lenzen} {et~al.}(1998){Lenzen}, {Hofmann}, {Bizenberger}, \&
  {Tusche}}]{1998SPIE.3354..606L}
{Lenzen}, R., {Hofmann}, R., {Bizenberger}, P., \& {Tusche}, A. 1998, in
  \procspie, Vol. 3354, Infrared Astronomical Instrumentation, ed. A.~M.
  {Fowler}, 606--614

\bibitem[{{Liu} {et~al.}(2013){Liu}, {Dupuy}, \&
  {Allers}}]{2013AN....334...85L}
{Liu}, M.~C., {Dupuy}, T.~J., \& {Allers}, K.~N. 2013, Astronomische
  Nachrichten, 334, 85

\bibitem[{{Macintosh} {et~al.}(2015){Macintosh}, {Graham}, {Barman}, {De Rosa},
  {Konopacky}, {Marley}, {Marois}, {Nielsen}, {Pueyo}, {Rajan}, {Rameau},
  {Saumon}, {Wang}, {Patience}, {Ammons}, {Arriaga}, {Artigau}, {Beckwith},
  {Brewster}, {Bruzzone}, {Bulger}, {Burningham}, {Burrows}, {Chen}, {Chiang},
  {Chilcote}, {Dawson}, {Dong}, {Doyon}, {Draper}, {Duch{\^e}ne}, {Esposito},
  {Fabrycky}, {Fitzgerald}, {Follette}, {Fortney}, {Gerard}, {Goodsell},
  {Greenbaum}, {Hibon}, {Hinkley}, {Cotten}, {Hung}, {Ingraham},
  {Johnson-Groh}, {Kalas}, {Lafreniere}, {Larkin}, {Lee}, {Line}, {Long},
  {Maire}, {Marchis}, {Matthews}, {Max}, {Metchev}, {Millar-Blanchaer},
  {Mittal}, {Morley}, {Morzinski}, {Murray-Clay}, {Oppenheimer}, {Palmer},
  {Patel}, {Perrin}, {Poyneer}, {Rafikov}, {Rantakyr{\"o}}, {Rice}, {Rojo},
  {Rudy}, {Ruffio}, {Ruiz}, {Sadakuni}, {Saddlemyer}, {Salama}, {Savransky},
  {Schneider}, {Sivaramakrishnan}, {Song}, {Soummer}, {Thomas}, {Vasisht},
  {Wallace}, {Ward-Duong}, {Wiktorowicz}, {Wolff}, \&
  {Zuckerman}}]{2015Sci...350...64M}
{Macintosh}, B., {Graham}, J.~R., {Barman}, T., {et~al.} 2015, Science, 350, 64

\bibitem[{{Maire} {et~al.}(2016{\natexlab{a}}){Maire}, {Bonnefoy}, {Ginski},
  {Vigan}, {Messina}, {Mesa}, {Galicher}, {Gratton}, {Desidera}, {Kopytova},
  {Millward}, {Thalmann}, {Claudi}, {Ehrenreich}, {Zurlo}, {Chauvin},
  {Antichi}, {Baruffolo}, {Bazzon}, {Beuzit}, {Blanchard}, {Boccaletti}, {de
  Boer}, {Carle}, {Cascone}, {Costille}, {De Caprio}, {Delboulb{\'e}},
  {Dohlen}, {Dominik}, {Feldt}, {Fusco}, {Girard}, {Giro}, {Gisler}, {Gluck},
  {Gry}, {Henning}, {Hubin}, {Hugot}, {Jaquet}, {Kasper}, {Lagrange},
  {Langlois}, {Le Mignant}, {Llored}, {Madec}, {Martinez}, {Mawet}, {Milli},
  {M{\"o}ller-Nilsson}, {Mouillet}, {Moulin}, {Moutou}, {Orign{\'e}}, {Pavlov},
  {Petit}, {Pragt}, {Puget}, {Ramos}, {Rochat}, {Roelfsema}, {Salasnich},
  {Sauvage}, {Schmid}, {Turatto}, {Udry}, {Vakili}, {Wahhaj}, {Weber}, \&
  {Wildi}}]{2016A&A...587A..56M}
{Maire}, A.-L., {Bonnefoy}, M., {Ginski}, C., {et~al.} 2016{\natexlab{a}},
  \aap, 587, A56

\bibitem[{{Maire} {et~al.}(2016{\natexlab{b}}){Maire}, {Langlois}, {Dohlen},
  {Lagrange}, {Gratton}, {Chauvin}, {Desidera}, {Girard}, {Milli}, {Vigan},
  {Zins}, {Delorme}, {Beuzit}, {Claudi}, {Feldt}, {Mouillet}, {Puget},
  {Turatto}, \& {Wildi}}]{2016SPIE.9908E..34M}
{Maire}, A.-L., {Langlois}, M., {Dohlen}, K., {et~al.} 2016{\natexlab{b}}, in
  \procspie, Vol. 9908, Ground-based and Airborne Instrumentation for Astronomy
  VI, 990834

\bibitem[{{Marley} {et~al.}(2012){Marley}, {Saumon}, {Cushing}, {Ackerman},
  {Fortney}, \& {Freedman}}]{2012ApJ...754..135M}
{Marley}, M.~S., {Saumon}, D., {Cushing}, M., {et~al.} 2012, \apj, 754, 135

\bibitem[{{Marois} {et~al.}(2014){Marois}, {Correia}, {V{\'e}ran}, \&
  {Currie}}]{2014IAUS..299...48M}
{Marois}, C., {Correia}, C., {V{\'e}ran}, J.-P., \& {Currie}, T. 2014, in IAU
  Symposium, Vol. 299, IAU Symposium, ed. M.~{Booth}, B.~C. {Matthews}, \&
  J.~R. {Graham}, 48--49

\bibitem[{{Marois} {et~al.}(2008){Marois}, {Macintosh}, {Barman}, {Zuckerman},
  {Song}, {Patience}, {Lafreni{\`e}re}, \& {Doyon}}]{2008Sci...322.1348M}
{Marois}, C., {Macintosh}, B., {Barman}, T., {et~al.} 2008, Science, 322, 1348

\bibitem[{{Marois} {et~al.}(2010){Marois}, {Zuckerman}, {Konopacky},
  {Macintosh}, \& {Barman}}]{2010Natur.468.1080M}
{Marois}, C., {Zuckerman}, B., {Konopacky}, Q.~M., {Macintosh}, B., \&
  {Barman}, T. 2010, \nat, 468, 1080

\bibitem[{{Mawet} {et~al.}(2015){Mawet}, {David}, {Bottom}, {Hinkley},
  {Stapelfeldt}, {Padgett}, {Mennesson}, {Serabyn}, {Morales}, \&
  {Kuhn}}]{2015ApJ...811..103M}
{Mawet}, D., {David}, T., {Bottom}, M., {et~al.} 2015, \apj, 811, 103

\bibitem[{{McLaughlin} {et~al.}(2006){McLaughlin}, {Anderson}, {Meylan},
  {Gebhardt}, {Pryor}, {Minniti}, \& {Phinney}}]{2006ApJS..166..249M}
{McLaughlin}, D.~E., {Anderson}, J., {Meylan}, G., {et~al.} 2006, \apjs, 166,
  249

\bibitem[{{Mesa} {et~al.}(2015){Mesa}, {Gratton}, {Zurlo}, {Vigan}, {Claudi},
  {Alberi}, {Antichi}, {Baruffolo}, {Beuzit}, {Boccaletti}, {Bonnefoy},
  {Costille}, {Desidera}, {Dohlen}, {Fantinel}, {Feldt}, {Fusco}, {Giro},
  {Henning}, {Kasper}, {Langlois}, {Maire}, {Martinez}, {Moeller-Nilsson},
  {Mouillet}, {Moutou}, {Pavlov}, {Puget}, {Salasnich}, {Sauvage}, {Sissa},
  {Turatto}, {Udry}, {Vakili}, {Waters}, \& {Wildi}}]{2015A&A...576A.121M}
{Mesa}, D., {Gratton}, R., {Zurlo}, A., {et~al.} 2015, \aap, 576, A121

\bibitem[{{Mesa} {et~al.}(2016){Mesa}, {Vigan}, {D'Orazi}, {Ginski},
  {Desidera}, {Bonnefoy}, {Gratton}, {Langlois}, {Marzari}, {Messina},
  {Antichi}, {Biller}, {Bonavita}, {Cascone}, {Chauvin}, {Claudi}, {Curtis},
  {Fantinel}, {Feldt}, {Garufi}, {Galicher}, {Henning}, {Incorvaia},
  {Lagrange}, {Millward}, {Perrot}, {Salasnich}, {Scuderi}, {Sissa}, {Wahhaj},
  \& {Zurlo}}]{2016A&A...593A.119M}
{Mesa}, D., {Vigan}, A., {D'Orazi}, V., {et~al.} 2016, \aap, 593, A119

\bibitem[{{Molli{\`e}re} {et~al.}(2017){Molli{\`e}re}, {van Boekel}, {Bouwman},
  {Henning}, {Lagage}, \& {Min}}]{2017A&A...600A..10M}
{Molli{\`e}re}, P., {van Boekel}, R., {Bouwman}, J., {et~al.} 2017, \aap, 600,
  A10

\bibitem[{{M{\"u}ller} {et~al.}(2018){M{\"u}ller}, {Keppler}, {Henning},
  {Samland}, {Chauvin}, {Beust}, {Maire}, {Molaverdikhani}, {vanBoekel},
  {Benisty}, {Boccaletti}, {Bonnefoy}, {Cantalloube}, {Charnay}, {Baudino},
  {Gennaro}, {Long}, {Cheetham}, {Desidera}, {Feldt}, {Fusco}, {Girard},
  {Gratton}, {Hagelberg}, {Janson}, {Lagrange}, {Langlois}, {Lazzoni}, {Ligi},
  {Menard}, {Mesa}, {Meyer}, {Molliere}, {Mordasini}, {Moulin}, {Pavlov},
  {Pawellek}, {Quanz}, {Ramos}, {Rouan}, {Sissa}, {Stadler}, {Vigan}, {Wahhaj},
  {Weber}, \& {Zurlo}}]{2018arXiv180611567M}
{M{\"u}ller}, A., {Keppler}, M., {Henning}, T., {et~al.} 2018, ArXiv e-prints
  [\eprint[arXiv]{1806.11567}]

\bibitem[{{Nielsen} {et~al.}(2008){Nielsen}, {Close}, {Biller}, {Masciadri}, \&
  {Lenzen}}]{2008ApJ...674..466N}
{Nielsen}, E.~L., {Close}, L.~M., {Biller}, B.~A., {Masciadri}, E., \&
  {Lenzen}, R. 2008, \apj, 674, 466

\bibitem[{{Pavlov} {et~al.}(2008){Pavlov}, {M{\"o}ller-Nilsson}, {Feldt},
  {Henning}, {Beuzit}, \& {Mouillet}}]{2008SPIE.7019E..39P}
{Pavlov}, A., {M{\"o}ller-Nilsson}, O., {Feldt}, M., {et~al.} 2008, in
  \procspie, Vol. 7019, Advanced Software and Control for Astronomy II, 701939

\bibitem[{{Rajan} {et~al.}(2017){Rajan}, {Rameau}, {De Rosa}, {Marley},
  {Graham}, {Macintosh}, {Marois}, {Morley}, {Patience}, {Pueyo}, {Saumon},
  {Ward-Duong}, {Ammons}, {Arriaga}, {Bailey}, {Barman}, {Bulger}, {Burrows},
  {Chilcote}, {Cotten}, {Czekala}, {Doyon}, {Duch{\^e}ne}, {Esposito},
  {Fitzgerald}, {Follette}, {Fortney}, {Goodsell}, {Greenbaum}, {Hibon},
  {Hung}, {Ingraham}, {Johnson-Groh}, {Kalas}, {Konopacky}, {Lafreni{\`e}re},
  {Larkin}, {Maire}, {Marchis}, {Metchev}, {Millar-Blanchaer}, {Morzinski},
  {Nielsen}, {Oppenheimer}, {Palmer}, {Patel}, {Perrin}, {Poyneer},
  {Rantakyr{\"o}}, {Ruffio}, {Savransky}, {Schneider}, {Sivaramakrishnan},
  {Song}, {Soummer}, {Thomas}, {Vasisht}, {Wallace}, {Wang}, {Wiktorowicz}, \&
  {Wolff}}]{2017AJ....154...10R}
{Rajan}, A., {Rameau}, J., {De Rosa}, R.~J., {et~al.} 2017, \aj, 154, 10

\bibitem[{{Rameau} {et~al.}(2013){Rameau}, {Chauvin}, {Lagrange}, {Boccaletti},
  {Quanz}, {Bonnefoy}, {Girard}, {Delorme}, {Desidera}, {Klahr}, {Mordasini},
  {Dumas}, \& {Bonavita}}]{2013ApJ...772L..15R}
{Rameau}, J., {Chauvin}, G., {Lagrange}, A.-M., {et~al.} 2013, \apjl, 772, L15

\bibitem[{{Rayner} {et~al.}(2009){Rayner}, {Cushing}, \&
  {Vacca}}]{2009ApJS..185..289R}
{Rayner}, J.~T., {Cushing}, M.~C., \& {Vacca}, W.~D. 2009, \apjs, 185, 289

\bibitem[{{Rousset} {et~al.}(2003){Rousset}, {Lacombe}, {Puget}, {Hubin},
  {Gendron}, {Fusco}, {Arsenault}, {Charton}, {Feautrier}, {Gigan}, {Kern},
  {Lagrange}, {Madec}, {Mouillet}, {Rabaud}, {Rabou}, {Stadler}, \&
  {Zins}}]{2003SPIE.4839..140R}
{Rousset}, G., {Lacombe}, F., {Puget}, P., {et~al.} 2003, in \procspie, Vol.
  4839, Adaptive Optical System Technologies II, ed. P.~L. {Wizinowich} \&
  D.~{Bonaccini}, 140--149

\bibitem[{{Samland} {et~al.}(2017){Samland}, {Molli{\`e}re}, {Bonnefoy},
  {Maire}, {Cantalloube}, {Cheetham}, {Mesa}, {Gratton}, {Biller}, {Wahhaj},
  {Bouwman}, {Brandner}, {Melnick}, {Carson}, {Janson}, {Henning}, {Homeier},
  {Mordasini}, {Langlois}, {Quanz}, {van Boekel}, {Zurlo}, {Schlieder},
  {Avenhaus}, {Beuzit}, {Boccaletti}, {Bonavita}, {Chauvin}, {Claudi}, {Cudel},
  {Desidera}, {Feldt}, {Fusco}, {Galicher}, {Kopytova}, {Lagrange}, {Le
  Coroller}, {Martinez}, {Moeller-Nilsson}, {Mouillet}, {Mugnier}, {Perrot},
  {Sevin}, {Sissa}, {Vigan}, \& {Weber}}]{2017A&A...603A..57S}
{Samland}, M., {Molli{\`e}re}, P., {Bonnefoy}, M., {et~al.} 2017, \aap, 603,
  A57

\bibitem[{{Sharp} \& {Burrows}(2007)}]{2007ApJS..168..140S}
{Sharp}, C.~M. \& {Burrows}, A. 2007, \apjs, 168, 140

\bibitem[{{Skemer} {et~al.}(2014){Skemer}, {Marley}, {Hinz}, {Morzinski},
  {Skrutskie}, {Leisenring}, {Close}, {Saumon}, {Bailey}, {Briguglio},
  {Defrere}, {Esposito}, {Follette}, {Hill}, {Males}, {Puglisi}, {Rodigas}, \&
  {Xompero}}]{2014ApJ...792...17S}
{Skemer}, A.~J., {Marley}, M.~S., {Hinz}, P.~M., {et~al.} 2014, \apj, 792, 17

\bibitem[{Skrutskie {et~al.}(2006)Skrutskie, Cutri, Stiening, Weinberg,
  Schneider, Carpenter, Beichman, Capps, Chester, Elias,
  {et~al.}}]{skrutskie2006two}
Skrutskie, M., Cutri, R., Stiening, R., {et~al.} 2006, \aj, 131, 1163

\bibitem[{{Soummer} {et~al.}(2012){Soummer}, {Pueyo}, \&
  {Larkin}}]{2012ApJ...755L..28S}
{Soummer}, R., {Pueyo}, L., \& {Larkin}, J. 2012, \apjl, 755, L28

\bibitem[{{Tuthill} {et~al.}(2000){Tuthill}, {Monnier}, {Danchi}, {Wishnow}, \&
  {Haniff}}]{tuthill2000keck}
{Tuthill}, P.~G., {Monnier}, J.~D., {Danchi}, W.~C., {Wishnow}, E.~H., \&
  {Haniff}, C.~A. 2000, \pasp, 112, 555

\bibitem[{{van Leeuwen}(2007)}]{2007AA...474..653V}
{van Leeuwen}, F. 2007, \aap, 474, 653

\bibitem[{{Vigan} {et~al.}(2010){Vigan}, {Moutou}, {Langlois}, {Allard},
  {Boccaletti}, {Carbillet}, {Mouillet}, \& {Smith}}]{2010MNRAS.407...71V}
{Vigan}, A., {Moutou}, C., {Langlois}, M., {et~al.} 2010, \mnras, 407, 71

\bibitem[{{Wright} {et~al.}(2010){Wright}, {Eisenhardt}, {Mainzer}, {Ressler},
  {Cutri}, {Jarrett}, {Kirkpatrick}, {Padgett}, {McMillan}, {Skrutskie},
  {Stanford}, {Cohen}, {Walker}, {Mather}, {Leisawitz}, {Gautier}, {McLean},
  {Benford}, {Lonsdale}, {Blain}, {Mendez}, {Irace}, {Duval}, {Liu}, {Royer},
  {Heinrichsen}, {Howard}, {Shannon}, {Kendall}, {Walsh}, {Larsen}, {Cardon},
  {Schick}, {Schwalm}, {Abid}, {Fabinsky}, {Naes}, \&
  {Tsai}}]{2010AJ....140.1868W}
{Wright}, E.~L., {Eisenhardt}, P.~R.~M., {Mainzer}, A.~K., {et~al.} 2010, \aj,
  140, 1868

\end{thebibliography}

\end{document}